\documentclass{article}
\usepackage[margin=2cm]{geometry}
\usepackage{graphicx,bm,amsmath,amssymb,latexsym,psfrag,epsfig,euscript, multirow,color,verbatim,cancel}
\title{Dark matter asymmetry through mass mixing}
\usepackage{mathrsfs}
\usepackage{float}

\allowdisplaybreaks
\begin{document}
\newcommand{\PL}{\mathcal{P}_{\mathrm L}}
\newcommand{\PR}{\mathcal{P}_{\mathrm R}}
\newcommand{\gs}{g_{\mathrm s}}
\newcommand{\be}{\begin{equation*}}
\newcommand{\ee}{\end{equation*}}
\newcommand{\ben}{\begin{equation}}
\newcommand{\een}{\end{equation}}
\newcommand{\bea}{\begin{eqnarray*}}
\newcommand{\eea}{\end{eqnarray*}}
\newcommand{\bean}{\begin{eqnarray}}
\newcommand{\eean}{\end{eqnarray}}
\newcommand{\MPl}{M_{\mathrm{Pl}}}
\newcommand{\Td}{T_{\mathrm{d}}}
\newcommand{\ptl}{\partial}

\begin{titlepage}

\begin{flushright}
\end{flushright}

\vspace{0.2cm}
\begin{center}
\Large\bf
Emergent Dark Matter, Baryon, and Lepton Numbers
\end{center}

\vspace{0.2cm}
\begin{center}
{\sc Yanou Cui, Lisa Randall and Brian Shuve}\\
\vspace{0.4cm}
{\sl Center for the Fundamental Laws of Nature\\
Jefferson Physical Laboratory\\
Harvard University\\
Cambridge, MA 02138, U.S.A.}
\end{center}

\vspace{0.2cm}
\begin{abstract}\vspace{0.2cm}
\noindent
We present a new mechanism for transferring a pre-existing lepton or baryon asymmetry to a dark matter asymmetry that relies on mass mixing which is dynamically induced in the early universe. Such mixing can succeed with only generic scales and operators and can give rise to distinctive relationships between the asymmetries in the two sectors. The mixing eliminates the need for the type of additional higher-dimensional operators that are inherent to many current asymmetric dark matter models.  We consider several implementations of this idea. In one model, mass mixing is temporarily induced during a two-stage electroweak phase transition in a two Higgs doublet model. In the other class of models, mass mixing is induced by large field vacuum expectation values at high temperatures - either moduli fields or even more generic kinetic terms.   Mass mixing models of this type can readily accommodate asymmetric dark matter masses ranging from 1 GeV to 100 TeV and expand the scope of possible relationships between the dark and visible sectors in such models.
\end{abstract}
\vfil

\end{titlepage}

\tableofcontents

\section{Introduction}

Asymmetric dark matter models \cite{Kaplan:2009ag} have the potential to explain the  fact that the energy in dark matter and ordinary matter are notably comparable. Suggested explanations involve the
co-generation of an asymmetry in both dark matter and baryonic sectors by the decay of heavy fields \cite{An:2009vq,Allahverdi:2010rh} or the Affleck-Dine mechanism \cite{Bell:2011tn}. Many others involve higher-dimensional operators that transfer asymmetries between the two, as was first explored in \cite{Nussinov:1985xr} and more recently in \cite{Kaplan:2009ag,Cohen:2009fz,Cohen:2010kn,Buckley:2010ui}. Although it seems reasonable that dark matter number and baryon or lepton number might not be exactly preserved in the early universe, most such models require new ingredients (scales or fields) to accommodate higher-dimensional operators that lead to requisite relationships between the dark matter and ordinary matter densities. Although conceptually compelling, these models leave open the question of whether we really expect the operators that violate baryon or lepton and dark matter numbers to exactly the right degree to really exist. Much more convincing would be models where there is mass mixing in the early universe arising from renormalizable interactions or operators suppressed only by the Planck scale. Ultimately, it is experiments that will hopefully shed light on the precise nature of the dark sector and its connection to the Standard Model.

In this paper, we show that it is possible that lepton (and hence baryon) number and dark matter number are not independent in the early universe. Mass mixing between dark-matter-number-carrying and lepton-number-carrying particles allows asymmetries to be established for both dark matter and lepton numbers. This mixing, of course, must turn off at later times when we know that the two symmetries must be independently preserved. The mechanism therefore requires fields whose value at high temperature differ from their low-temperature values so that in the late universe, dark matter and lepton numbers are independent symmetries and separately conserved.

The mass parameters in the models we present are linked to scales we know should be associated with new fields or interactions, namely the Planck scale and the electroweak scale (although in some cases we invoke also leptogenesis or grand unification scales). Furthermore, mass mixing can allow for novel relationships between the asymmetries in the dark and visible sectors when asymmetry transfer processes are out of equilibrium or the mass mixing shuts off rapidly in a first-order phase transition, allowing for deviations from the often-cited ``predicted'' value of the dark matter mass of $\sim5$ GeV.

In our models, the mass mixing is active during a period of time after a fundamental $B-L$ asymmetry is generated but vanishes at late times to ensure stability of the dark matter. This is the opposite of the usual chronology of phase transitions, where finite-temperature corrections keep the system in the symmetric phase at early times and a vacuum expectation value (VEV) arises at later times. There are at least two situations when  symmetries  are broken in the early universe and restored at late times. One involves multiple fields where one field initially stabilizes another field away from its minimum but no longer does so below a critical value \cite{Weinberg:1974hy}. Two stage phase transitions in a two Higgs model were first discussed in \cite{Land:1992sm}. A similar idea is exploited in hybrid inflation models, \cite{Linde:1991km}, but  the fields in our models are confined to the minima of the potential and involve no slow roll. We present models with multiple-stage phase transitions in Section \ref{sec:twost}. The second possibility is temperature-dependent background fields in the early universe that turn off as the universe cools. Examples include moduli field VEVs and the background energy density in the early universe. When these background fields are active, Planck-suppressed higher-dimensional operators can induce mass mixing at early times.  We present both supersymmetric and non-supersymmetric models in Section \ref{sec:planck}.

\section{Two-stage phase transition} \label{sec:twost}
\subsection{Overview}
Canonical examples of phase transitions in the early universe involve a transition from the symmetric phase at high temperatures to a broken phase at low temperatures.  In Weinberg's pioneering work on symmetry breaking at finite temperature \cite{Weinberg:1974hy}, examples were presented where the reverse is true: a  symmetry broken at high temperatures is restored at later times. Typically, there exist multiple scalar fields and multiple stages to such phase transitions.  In this paper, we first discuss one simple example of a two Higgs doublet model originally considered in the context of electroweak baryogenesis \cite{Land:1992sm}, although our results can be generalized in a relatively straightforward manner. The models in the following section are more generic, but have slightly more technically complicated dynamics and work for a more restricted parameter range.

In the two Higgs set-up, a $B-L$ asymmetry is established at temperatures above a few hundred GeV. The precise mechanism is irrelevant to our analysis. At high temperatures $T\gg 100$ GeV, the universe is in an electroweak symmetric phase. As the universe cools to $T\sim 100$ GeV,  the electroweak phase transition occurs, proceeding in two stages. In the first stage, a non-Standard Model scalar $\Phi$ acquires a VEV while the Higgs remains confined to the origin. When $\Phi$ gets a VEV, it induces mass mixing between the dark sector and the left-handed Standard Model leptons. During this period, the lepton asymmetry gets shifted into an asymmetry involving the high-temperature mass eigenstates, which are linear combinations of the zero-temperature dark matter and lepton fields.

At later time, a minimum develops along the Standard Model Higgs $H$ direction with a lower energy than the initial vacuum. In the new vacuum, the Standard Model Higgs develops a VEV but $\Phi$ is confined to the origin by the potential induced by the large Standard Model Higgs VEV. Bubbles of the new vacuum form in a first-order phase transition. Because this phase transition occurs very rapidly,  mass mixing shuts off before thermal interactions can adjust the asymmetry. This allows for non-trivial mixing of the asymmetry between the dark and visible sectors. In one scenario, the mixing can populate heavy dark sector fields that have been thermally suppressed in the previous stage, in which case the relationship between the final dark matter and lepton asymmetries is determined by the mixing angle at the time of the first-order phase transition. In another scenario, it is determined purely by thermal factors.

In this two Higgs model, the  second Higgs plays two important roles. First, it allows a field value that turns on and off before the fields enter the true electroweak-symmetry-breaking vacuum with the zero temperature Higgs VEV. Second,  the  $\Phi$ scalar is initially trapped far from the origin and allows for a first-order phase transition so the asymmetry in the dark matter gets established immediately.

We propose models with $\mathcal O(1)$ couplings and mass scales in the theory that are connected to the weak scale. We do not attempt to solve the hierarchy problem, but presume the model can be embedded in models that do have solutions (for instance, in supersymmetric models in which  both scalars get their masses from supersymmetry breaking).

\subsection{Asymmetry transfer during phase transition}\label{sec:asymtransfer}
The  mechanism of asymmetry transfer by mass mixing is largely independent of the specific details of the two Higgs model, and we postpone a discussion of the field content, interactions, and phase transition to Section \ref{sec:twohiggsfields} and appendices \ref{app:twohiggs} and \ref{app:bubble}. For now, we consider general models with a doublet scalar $\Phi$ with a Yukawa coupling $y_X$ to a vector-like dark matter particle, $X$, and a left-handed lepton $L$. During the intermediate stage of the electroweak phase transition, the real part of the scalar $\Phi$ has a VEV that induces mass mixing between $X$ and $L$,
\ben \label{eq:massmix}
m_{XL}(T) = y_X\,v_{\Phi}(T).
\een
If $\Phi$ does not acquire a VEV, an unbroken $\mathrm U(1)_{X-\Phi}$ symmetry would persist through the entire universe's evolution. At late times, after $\Phi$ had decayed away entirely, the solution to the chemical potential relations would be $\mu_{\Phi}=\mu_X=0$, and there would be no late time $X$ asymmetry. Thus, the mass mixing stage is vital to ensuring a non-zero asymmetry at  late times.

Although the model contains three generations of $X$ and $L$, for simplicity and to retain only the essential elements of the mass mixing transfer mechanism, we consider only the case where one flavor of $L$ and $X$ have a large mixing. This could be obtained, for example, by hierarchical Yukawa couplings of the same form as in the Standard Model. The discussion can be easily generalized to the case of arbitrary mixings between flavor states.

The vector-like dark matter has a Dirac mass $m_X$. The mass matrix in the $(X,\bar X,L^0)$ basis, including the dynamically induced mixing term $m_{XL}(T)$ is
\ben
\mathcal M = \left(\begin{array}{ccc} 0 & m_X & m_{XL}(T) \\
m_X & 0 & 0 \\
m_{XL}(T) & 0 & 0 \end{array}\right),
\een
which gives two degenerate states with mass $M = \sqrt{m_X^2+m_{XL}^2}$ and a massless state. The mass basis states are
\bean
|L'\rangle &=& -\sin\theta |\bar X\rangle  + \cos\theta |L^0\rangle,\\
|\bar X'\rangle &=& \cos\theta |\bar X\rangle +\sin\theta|L^0\rangle,\\
|X\rangle,
\eean
with the $|L'\rangle$ state being the massless state and the $|X\rangle$, $|\bar X\rangle$ states having mass $M$. The mixing angle is
\ben\label{eq:sine}
\sin\theta = \frac{m_{XL}}{\sqrt{m_X^2+m_{XL}^2}}.
\een

In the interaction basis, the rate of asymmetry transfer between $L$ and $X$ is approximately    \cite{Manohar:1986gj}
\ben\label{eq:transferrate}
\Gamma_{L\rightarrow X} = \Gamma_0\sin^22\theta\sin^2\left(\frac{M^2}{6T\Gamma_0}\right),
\een
which is explained in Section \ref{sec:massmixing}. This formula  says that the net transfer rate is the oscillation rate that occurs before a thermal interaction, multiplied by the thermal interaction rate. $\Gamma_0$ is the rate at which scattering events occur, averaged between $X$ and $L$. The system is in chemical equilibrium for the mass scales and couplings considered in this section if $y_X\gtrsim10^{-6}$.

The gauge interaction of $L^0$ with charged leptons appears in the mass basis as
\ben
\mathcal L_{\mathrm{gauge}}= \frac{1}{2}(L^-)^{\dagger}\gamma^{\mu}(1-\gamma^5)W^-_{\mu}\left(\cos\theta\,L' + \sin\theta \bar X'\right).
\een
The rate of interactions of $\bar X'$ is $\sin^2\theta$ times the usual weak interaction rate. Since the Hubble constant is very small at the time of the electroweak phase transition, for any non-negligible value of $\theta$, $\bar X'$ is in thermal equilibrium due to gauge interactions. Since $L^-$ carries lepton number of +1,   $\bar X'$ and $L'$ carry lepton number of +1 as well. The Yukawa interaction
\ben
\mathcal L \supset y_X\,\Phi\,XL = y_X\,\Phi\,X(-\sin\theta \bar X'+\cos\theta L')
\een
gives $X$ a lepton number $-1$. The interactions in our model are sufficient to keep the dark matter in thermal and chemical equilibrium.

We define the asymmetries in each field as $\Delta n_{L'}$, $\Delta n_{\bar X'}$ and $\Delta n_X$. The asymmetries in the fields are (for $T < M$)
\bean
\Delta n_{L'} &=& \frac{1}{3}\mu_L\, T^2,\label{eq:asymn}\\
\Delta n_{\bar X'} = -\Delta n_{X} &=&\sqrt{\frac{2TM^3}{\pi^3}}\,\mu_L\, e^{-M/T}\,\,\,\,\,\,\,\,(M\gg T),\label{eq:asymx}\\
&=& \frac{1}{3}\mu_L\,T^2\,\,\,\,\,\,\,\,\,\,\,\,\,\,\,\,\,\,\,\,\,\,\,\,\,\,\,\,\,\,\,\,\,\,\,\,\,\,\,\,(M\ll T).
\eean
where $\mu_L$ is the chemical potential for lepton number.  $\mu_L$ is fixed  because the baryon and lepton numbers have already equilibriated due to sphaleron processes active prior to the electroweak phase transition. At the time of the second stage of the phase transition, the weak gauge boson masses are large enough that sphaleron processes are exponentially suppressed.

These relations are valid as long as the mixing is turned on. At a temperature $T_{\mathrm N}$, bubble nucleation occurs and the system moves to the true vacuum via a strongly first-order phase transition. The dynamics of the phase transition are discussed in Appendix \ref{app:bubble}. Therefore, in the moment immediately following the phase transition, the particles are still in the same state that they were in prior to the phase transition (according to the instantaneous approximation), but their masses are now fixed by the new vacuum. In particular, the mass mixing has shut off. The abundances of $X$, $\bar X$, and $L^0$ are fixed by the abundances of the mass eigenstates at decoupling. In particular,
\bean
\Delta n_{L} &=& \cos^2\theta \Delta n_{L'} + \sin^2\theta \Delta n_{\bar X'}, \\
\Delta n_{\bar X} &=& \sin^2\theta\Delta n_{L'} + \cos^2\theta\Delta n_{\bar X'},\\
\Delta n_{X} &=&-\Delta n_{\bar X'}
\eean
We now define the total dark matter and baryon numbers (obtained from summing over all charged fields) as
\bean
\Delta_X &=& \Delta n_X - \Delta n_{\bar X},\\
\Delta_B &=& \Delta n_B-\Delta n_{\bar B}.
\eean
The ratio of $X$ number to $\Delta n_{L}$ is
\ben\label{eq:asymxnratio}
\frac{\Delta_X}{\Delta n_{L}} = -\frac{(1+\cos^2\theta)\Delta n_{\bar X'}+\sin^2\theta\Delta n_{L'}}{\cos^2\theta\Delta  n_{L'}+\sin^2\theta \Delta n_{\bar X'}}.
\een
Once the phase transition completes, there is a conserved global $\mathrm U(1)_{X-\Phi}$ symmetry. Since $\Delta n_{\Phi}=0$ at the time of bubble nucleation, and $\Phi$ decays to $X$ at late times, the zero temperature value of $\Delta_X$ is identical to that immediately after the phase transition occurs.

 Depending on parameters, the ratio of the $X$ and $L$ asymmetries in (\ref{eq:asymxnratio}) can be consistent with both light dark matter with a mass of $\sim$ GeV, or with heavier dark matter masses in the range $200-500$ GeV (other models consistent with heavier asymmetric dark matter can be  found in \cite{Buckley:2010ui,Cohen:2009fz,Allahverdi:2010rh}). For the heavy case, the final asymmetry ratio is dependent on which terms dominate (\ref{eq:asymxnratio}).  We identify three distinct limits:
\begin{enumerate}
\item {\bf Relativistic $X$}. In the relativistic limit, when $T_{\mathrm N}\gg m_X$, $m_{XL}$, there is no thermal suppression in the $X_1$ distribution and we have $\Delta n_{L'} =\Delta n_{\bar X'}$. Equation (\ref{eq:asymxnratio}) reduces to
\ben\label{eq:rellimit}
\frac{\Delta_X}{\Delta n_{L}}\approx -\frac{2}{3}.
\een
This is consistent with light dark matter, $m_X\approx 3.3$ GeV, which we calculate from (\ref{eq:rellimit2}) below and by imposing $\Omega_X =5\Omega_{\rm B}$.

\item {\bf Thermally-suppressed $X$}. In the non-relativistic limit $m_X\gg T_{\mathrm N}$, then $\Delta n_{\bar X'}\ll\Delta n_{L'}$.  In the thermal-suppression-dominated limit, $\Delta n_{\bar X'}/\Delta n_{L'}\gg\tan^2\theta$ and
\ben\label{eq:thermallimit}
\frac{\Delta_X}{\Delta n_{L}} \approx -6\sqrt{\frac{2M^3}{\pi^3T^3}}\,e^{-M/T}.
\een
This relation is the same as the one appropriate to asymmetry transfer via higher-dimensional operators that freeze out when $X$ is non-relativistic  \cite{Buckley:2010ui}. This is the relevant limit when the $\phi$ VEV at $T_{\mathrm N}$ or the Yukawa coupling $y_X$ is small.

\item {\bf Mixing-angle-suppressed $X$}. The other possible non-relativistic solution is   when $\Delta n_{\bar X'}/\Delta n_{L'}\ll\tan^2\theta$ and the solution is mixing-angle-dominated. In this case, the final ratio of $X/L$ asymmetries is independent of $\Delta n_{\bar X'}$ and is determined only by the mixing angle,
\ben\label{eq:anglelimit}
\frac{\Delta_X}{\Delta n_{L}} = -\tan^2\theta.
\een
This is a novel relationship and unique to models with mass mixing. Unlike the thermal-suppression-dominated limit, this result is only weakly dependent on $T_{\mathrm N}$, since the expression for the $\phi$ VEV (\ref{eq:phivev}) is effectively temperature independent at low $T$.

\end{enumerate}
We show below that, when the mixing is large ($y_X\gtrsim1$), both the mixing angle and thermal suppression terms in (\ref{eq:asymxnratio}) are relevant.

We now relate the baryon number to the $X$ number using the relationship between baryon and lepton numbers in the early universe \cite{Harvey:1990qw},
\ben
\frac{\Delta_B}{\Delta_L} = -\frac{4}{3}\,\frac{6N_{\rm f}+3}{14N_{\rm f}+9},
\een
where $N_{\rm f}$ is the number of Standard Model flavors. The baryon number to $X$ number in each of the above cases is
\begin{enumerate}

\item \ben\frac{\Delta_X}{\Delta_{\rm B}} = \frac{2}{3},\label{eq:rellimit2}\een

\item \ben\frac{\Delta_X} {\Delta_{\rm B}}= \frac{3}{\frac{2}{3}\sqrt{\frac{\pi^3T^3}{2M^3}}\,e^{M/T}-4},\label{eq:thermalresult}\een

\item\ben\label{eq:xbfrac}
\frac{\Delta_X}{\Delta_{\rm B}} =\frac{3\tan^2\theta}{4(1-\tan^2\theta)}.
\een

\end{enumerate}

\subsection{Field content and interactions}\label{sec:twohiggsfields}
We now describe our model in more detail. It contains:
\begin{itemize}

\item The Standard Model fields, including and most relevant, the Higgs $H$ and left-handed leptons $L_i$.

\item The dark matter fields, which are three vector-like fermions $X_i$ carrying a lepton flavor index.

\item An additional doublet scalar $\Phi$ with hypercharge $+1/2$. The mass of $\Phi$ satisfies $m_{\Phi} > m_X$ so that $X$ is stable.

\end{itemize}
There exists a $Z_2$ symmetry, under which $\Phi$ and $X$ are charged but Standard Model fields are not. This makes $X$ absolutely stable at late times, since $\langle\Phi\rangle=0$ in the true vacuum and the $Z_2$ symmetry is unbroken. The symmetry also excludes the term $H\Phi^\dagger$, which would give $\Phi$ a tadpole when $H$ gets a VEV.

The new terms of the Lagrangian are
\ben
\mathcal L \supset m_X X_i\bar X_i + y_X\,\Phi X_i L_i + V(H,\Phi) +\mathrm{h.c.},
\een
where
\ben\label{eq:higgspotential}
V(T=0) = 4k_1|H|^4-4\mu_1^2|H|^2+4k_2|\Phi|^4-4\mu_2^2|\Phi|^2 + 4k_3|\Phi|^2|H|^2,
\een
using the notation of \cite{Land:1992sm} that is convenient in the basis of real scalar fields. In addition to ensuring dark matter stability, the $Z_2$ symmetry excludes an $H\Phi^{\dagger}$ term, which would give $\Phi$ a tadpole  when $H$ gets a VEV.

With the convention in (\ref{eq:higgspotential}), perturbativity requires that $k_1,k_2<0.8$ and $k_3<3.1$. There are additional constraints to avoid hitting a Landau pole below the GUT scale, but this is highly dependent on what other fields couple to $H$ and $\Phi$ (for example, the quartic terms could have large contributions from integrating out other weak scale particles) so we do not consider this as a constraint.

At $T=0$, $\Phi$ must be stabilized at the origin, while $H$ should have its measured VEV of $v=246$ GeV. The vacua of the theory are determined by finding the critical points of the potential. To do so, it is simplest to move to the basis of real fields. Because $\mathrm U(1)_{\mathrm{em}}$ remains unbroken, the VEVs can be rotated purely into the neutral components of $H$ and $\Phi$. Furthermore, assuming that the potential is $CP$-conserving (for simplicity), only the real components of $\Phi$ and $H$ can get VEVs. Therefore, we make the substitution
\ben
\Phi = \frac{1}{\sqrt 2}\left(\begin{array}{c} \phi \\ 0 \end{array}\right),\,\,\,\,\,\,\, H=\frac{1}{\sqrt 2}\left(\begin{array}{c} h \\ 0 \end{array}\right),
\een
giving the zero temperature potential
\ben
V(T=0) = k_1 h^4-2\mu_1^2 h^2+k_2\phi^4-2\mu_2^2\phi^2+k_3\phi^2 h^2.
\een

Because we are interested in the evolution of the field values as the universe cools, we also need to include thermal contributions to the masses of $H$ and $\Phi$. These thermal corrections are proportional to $T$ and dominate over the tree-level masses for $T\gg\mu$. Ultimately, these corrections   determine how the phase transition  proceeds. We can write the finite temperature potential as
\ben
V(T) = k_1 h^4-2\mu_1^2 h^2+k_2\phi^4-2\mu_2^2\phi^2+k_3\phi^2 h^2 + \frac{1}{2}\alpha_1T^2 h^2+\frac{1}{2}\alpha_2 T^2 \phi^2,
\een
where
\bean
\alpha_1 &=& 2k_1+\frac{2}{3}k_3 + \frac{e^2(1+2\cos^2\theta_{\mathrm W})}{\sin^2 2\theta_{\mathrm W}}+\frac{1}{2}y_t^2,\\
\alpha_2 &=& 2k_2+\frac{2}{3}k_3 + \frac{e^2(1+2\cos^2\theta_{\mathrm W})}{\sin^2 2\theta_{\mathrm W}}+\frac{1}{6}y_X^2.
\eean
The light quark and lepton Yukawa couplings are sufficiently small that only the top Yukawa needs to be considered for corrections to the $h$ mass.

We give the details of the vacua and constraints on the parameters in Appendix \ref{app:twohiggs}. The first of the two most important results is the constraint leading to a two stage phase transition,
\ben \label{eq:constraint}
\sqrt{\frac{k_1}{k_2}} < \frac{\mu_1^2}{\mu_2^2} < \frac{\alpha_1}{\alpha_2},
\een
which comes from requiring that the $\phi$ direction become unstable at the origin before $h$, while also requiring that the Standard Model Higgs vacuum have lower energy than the $\phi\neq0$ vacuum. The other main requirement is that $\phi$ be stabilized to the origin in the true vacuum,
\ben\label{eq:phimass}
m_{\phi}^2 = 2\mu_1^2\frac{k_3}{k_1}-4\mu_2^2>0.
\een

\subsection{Numerical results}\label{sec:numerics}

We present numerical results for a scan of parameters leading to a two stage phase transition. The most relevant result for us  is the temperature $T_{\mathrm N}$ at which bubble nucleation occurs for a given set of parameters. This determines which of the three limits (relativistic, thermal-suppression-dominated or mixing-angle-dominated) is most relevant and hence the dark matter mass consistent with $\Omega_{\mathrm{DM}} = 5\,\Omega_{\mathrm{baryon}}$. We present details of the calculation of $T_{\rm N}$ in Appendix \ref{app:bubble}.
\begin{figure}[t]
\begin{center}
\includegraphics[scale=0.5]{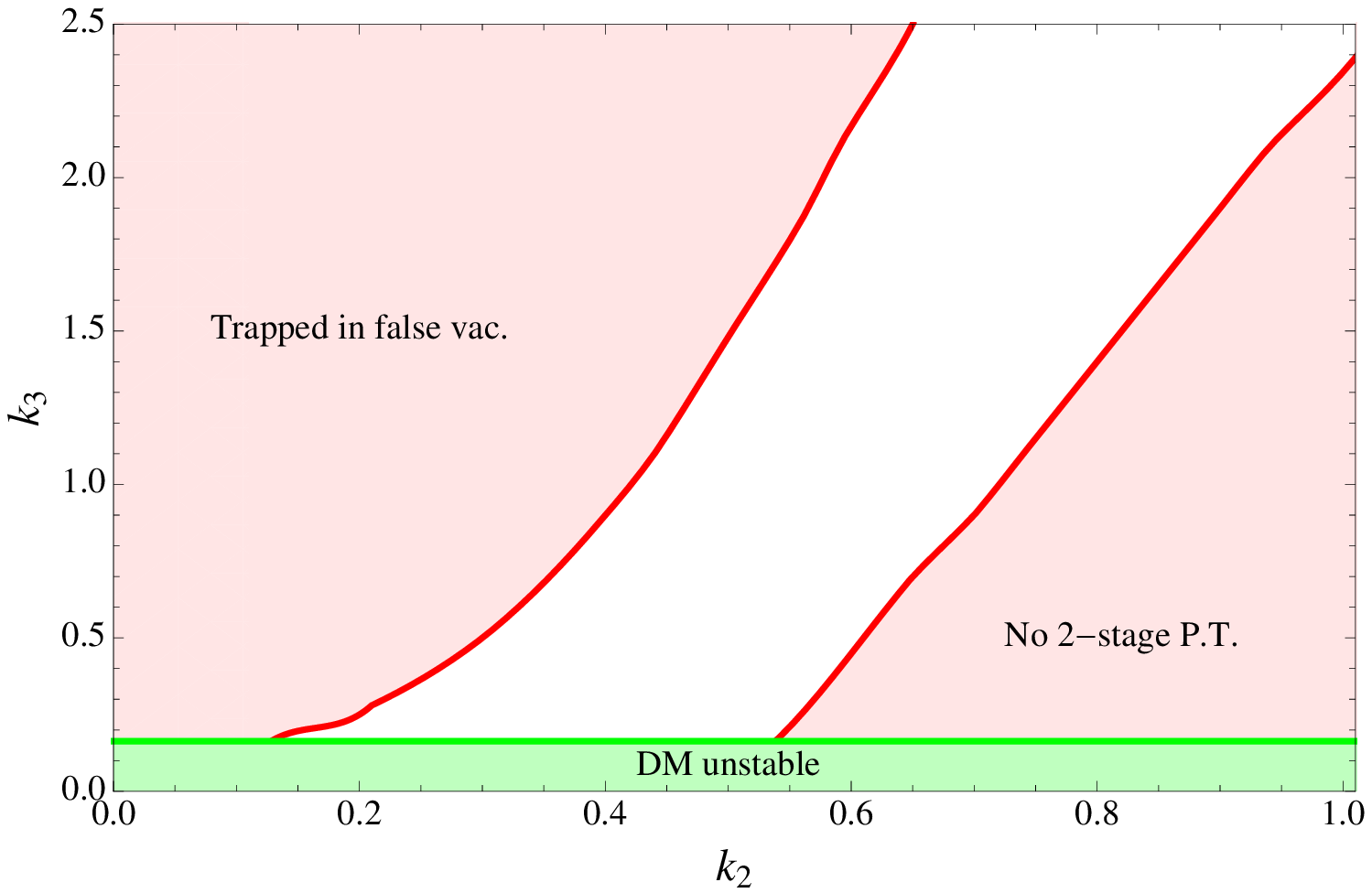}
\includegraphics[scale=0.5]{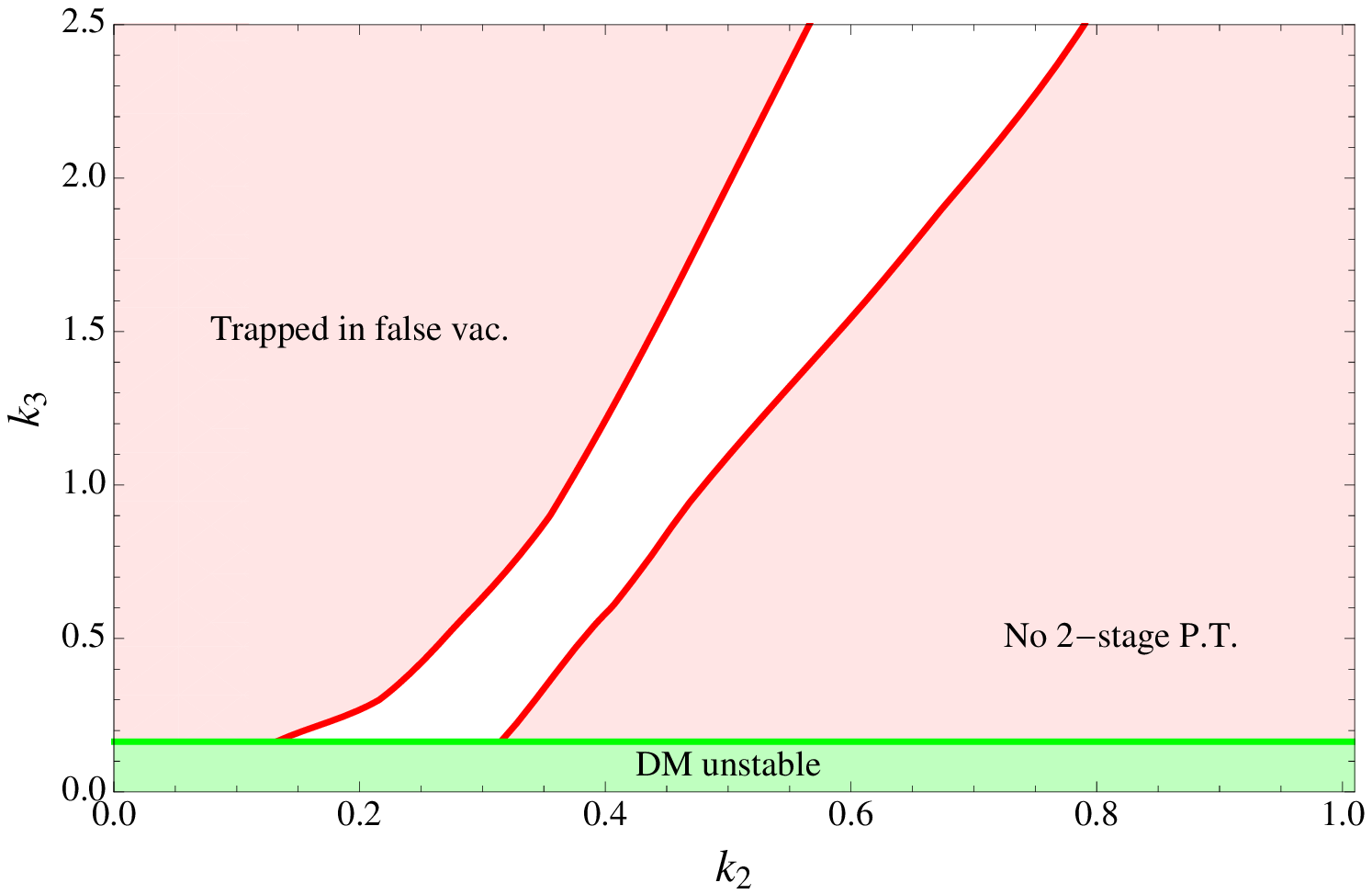}
\caption{Unshaded regions in the $k_2-k_3$ plane give rise to a two stage phase transition and a viable dark matter asymmetry through mass mixing for light dark matter. Each parameter point is consistent with dark matter of mass $3.3$ GeV. The other parameters are held fixed at $\mu_2=54$ GeV for both plots, $y_X=0.5$ (left) and $y_X=1.7$ (right). In the left shaded region (red), the barrier between vacua is too large for rapid bubble nucleation and the system remains trapped in the false vacuum. In the right shaded region (also red), the $\phi\neq0$ direction is not a stable vacuum. In the bottom shaded region (green), the scalar mass $m_\phi$ is excluded by LEP.}
\label{fig:TwoStage}
\end{center}
\end{figure}

The necessary and sufficient conditions for asymmetry transfer through a two stage phase transition are:
\begin{itemize}

\item The constraints (\ref{eq:constraint}) are satisfied.

\item The bubble nucleation temperature $T_{\mathrm N}>0$, and the wall velocity is large enough that the transition is strongly first-order (see Appendix \ref{app:bubble}).

\item The mass $m_{\phi} > m_X$  in the final vacuum so that $X$ is stable, where $m_\phi$ is given in (\ref{eq:phimass}).

\end{itemize}
There is also a constraint coming from LEP searches for new doublet scalars (specifically sleptons, from which there is a bound $m_{\phi}\gtrsim90$ GeV \cite{Abbiendi:2003ji}).

The Standard Model Higgs mass parameter, $\mu_1$, is fixed to be $\mu_1=42$ GeV by a Higgs boson mass of 120 GeV in the final vacuum, while the quartic coupling $k_1$ is fixed by the VEV $\langle h\rangle=246$ GeV.    We have a four-dimensional parameter space, consisting of the $\phi$ quartic self-coupling $k_2$, the $\phi-h$ mixed quartic coupling $k_3$, the $\phi$ mass parameter $\mu_2$, and the Yukawa coupling $y_X$.

The allowed values of $\mu_2$ come from the constraint (\ref{eq:constraint}). The range is $\mu_2=25-100$ GeV, which is precisely the range we expect if $\mu_1$ and $\mu_2$ have a common origin. Since perturbativity limits $k_2\lesssim0.8$,  the upper bound of $\mu_2<100$ GeV follows from (\ref{eq:constraint}). The lower bound $\mu_2>25$ GeV comes from the fact that the ratio $\alpha_1/\alpha_2$ is fixed by the Standard Model interactions when $y_X,k_2,k_3\rightarrow0$. The parameter space for a viable two stage phase transition is small close to either bound on $\mu_2$, so we present all of our results with an intermediate, representative value of $\mu_2=54$ GeV and scan over other parameters.

The Yukawa coupling is selected to be $\mathcal O(1)$. In this case, the symmetric component of $X$ annihilates through the Yukawa interaction (see Section \ref{sec:annihilation}).

\begin{figure}[t]
\begin{center}
\includegraphics[scale=0.5]{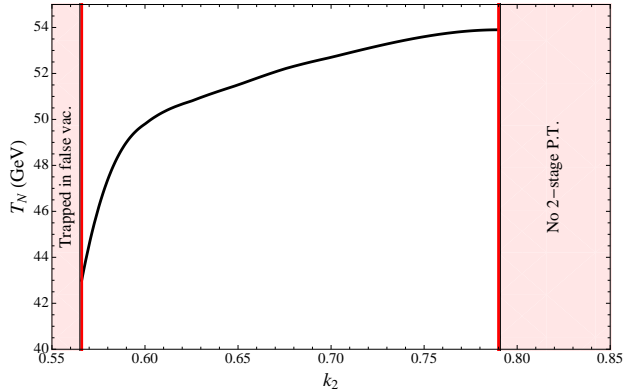}
\caption{Relationship between the temperature of the first-order phase transition $T_{\rm N}$ and the $\phi$ quartic coupling $k_2$. The other parameters are held fixed at $k_3=1.5$, $y_X=1.7$ and $\mu_2=54$ GeV. No first-order phase transition takes place in the shaded regions.}
\label{fig:TunnellingT}
\end{center}
\end{figure}

We now present, in turn, the parameters consistent with the three dark matter scenarios discussed above.
\begin{enumerate}
\item {\bf Relativistic $X$}. In general, $T_{\mathrm N}\gg3.3$ GeV, so  light $3.3$ GeV dark matter is consistent with any choice of parameters leading to a two stage phase transition. We show    in Figure \ref{fig:TwoStage} a region of parameter space in the $k_2-k_3$ plane (with $\mu_2=54$ GeV and two values of $y_X=0.5,1.7$) with a two stage phase transition and light dark matter.

In the right shaded region, the couplings $k_2$ give a large contribution to the thermal mass of $\phi$, stabilizing the $\phi$ direction and leading the system to condense directly into the true, $h\neq0$ vacuum. In this case, the $\phi$ VEV is never non-zero. For smaller values of $k_2$, the VEV $\langle\phi\rangle \approx \mu_2/\sqrt{k_2}$ increases and the barrier between vacua becomes larger. In the left shaded region, the barrier is so large that bubble nucleation never occurs and the system remains stuck in the false vacuum. The tunneling temperatures are shown in Figure \ref{fig:TunnellingT}.

In the shaded region at the bottom, $m_\phi$ from (\ref{eq:phimass}) is smaller than the LEP bound.

\item {\bf Thermally-suppressed $X$}. The relation (\ref{eq:thermalresult}) dictates the $X$ masses consistent with a given tunneling temperature $T_{\mathrm N}$. We present our results in Figure \ref{fig:Thermal} as a function of the parameters $k_2$, $k_3$, with $\mu_2=54$ GeV and $y_X=0.5$. The Yukawa coupling is small enough that the mixing angle contribution to the asymmetry is always smaller than the thermal terms. For Yukawa couplings $y_X\gtrsim1$, both mixing angle and  thermal suppressions are relevant and we address this case in point \#3.

 The entire unshaded region is consistent with dark matter with a  thermally-suppressed asymmetry, and the dashed contours illustrate a few representative values of $m_X$ between 300 and 450 GeV. The outer side shaded regions represent the same constraints as  the relativistic case. For small values of $k_3$, $m_\phi$ becomes smaller than the value of $m_X$ giving the correct dark matter density, making the dark matter unstable. In the inner shaded region on the left, the mixing angle contribution to $\Delta_X/\Delta_{\rm B}$ dominates over the thermally-suppressed terms.

We find a range of dark matter masses $300-500$ GeV for the given parameters. When we scan over other values of $\mu_2$, we find a range $\sim250-550$ GeV, although most of the parameter space lies in the $300-450$ range. According to (\ref{eq:constraint}), a smaller value of $\mu_2$ gives a two stage phase transition with smaller values of $k_2$, shifting the band in Figure \ref{fig:Thermal} to the left (to smaller values of $k_2$). Smaller values of $y_X$ decrease the value of $\alpha_2$, allowing larger values of $k_2$ and relaxing the upper bound on $k_2$ from (\ref{eq:constraint}).

\begin{figure}[t]
\begin{center}
\includegraphics[scale=0.5]{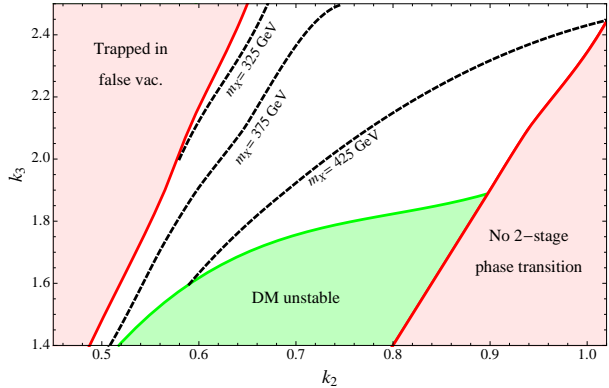}
\caption{Regions in the $k_2-k_3$ plane where the final $X$ asymmetry is thermally suppressed. The unshaded region is the viable parameter space. Other parameters are held fixed at $\mu_2=54$ GeV and $y_X=0.5$. Dashed contours show representative $X$ masses associated with each point in parameter space (from left to right, 325 GeV, 375 GeV, 425 GeV). The outer shaded regions to the side (red) do not give rise to a two stage phase transition, while the shaded bottom region (green) violates the condition $m_X<m_{\phi}$, rendering $X$ unstable.}
\label{fig:Thermal}
\end{center}
\end{figure}

\begin{figure}[h]
\begin{center}
\includegraphics[scale=0.7]{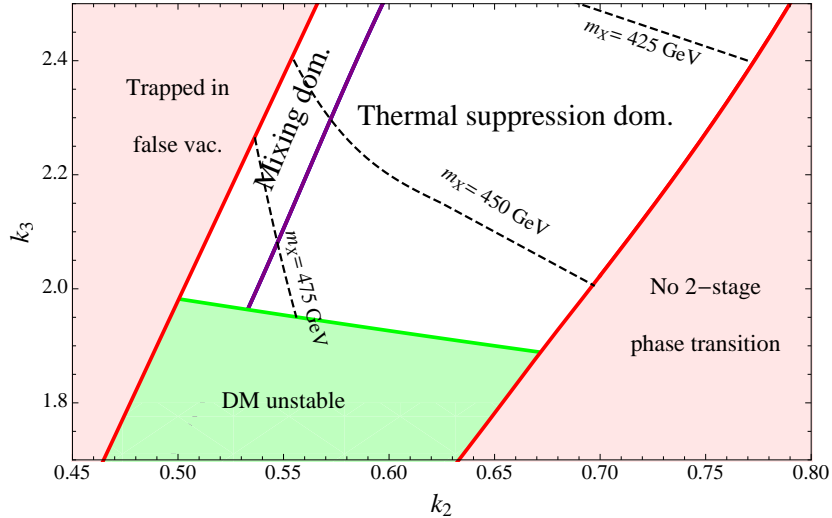}
\caption{Regions in the $k_2-k_3$ plane with $y_X=1.7$ and $\mu_2=54$ GeV consistent with weak scale dark matter. The mixing angle contribution is relevant for all  points. The unshaded region is the viable parameter space. Dashed contours show representative $X$ masses associated with each parameter  (from left to right, 425 GeV, 450 GeV, 475 GeV). The two inner, unshaded regions show parameters where the mixing angle (left) and thermal suppression (right) are respectively dominant. The outer shaded regions (red) do not give rise to a two stage phase transition. In the lower shaded region (green), the $m_X<m_{\phi}$ condition is violated.}
\label{fig:Angle}
\end{center}
\end{figure}

 This range of masses is the expected range for the two Higgs model since  the bubble nucleation temperature $T_{\mathrm N}\lesssim2\mu_1$ from (\ref{eq:degenerate}), and this restricts  $m_{\phi}\lesssim550$ GeV. For stability, the $X$ mass can never exceed the $\phi$ mass, and the largest possible value of $m_{\phi}$ with perturbative quartic couplings is about 550 GeV. The lower bound on masses comes from the fact that the vast majority of points in parameter space with a two stage phase transition have tunneling temperature $T_{\rm N} \gtrsim35$ GeV (when the vacua become degenerate at a temperature lower than this, the energy barrier is typically large enough that the system remains trapped in the false vacuum). Thus, when the final asymmetry in $X$ is determined by thermal suppression, the range of allowed dark matter masses is $250\lesssim m_X\lesssim550$ GeV.

\item {\bf Mixing-angle-suppressed $X$}.  We present our results in Figure \ref{fig:Angle} as a function of the parameters $k_2$, $k_3$, with $\mu_2=54$ GeV and  $y_X=1.7$\footnote{For $y_X=0.5$ considered in the first two cases, the mixing angle is always too small to give a significant contribution to the asymmetry.}.  Dark matter is consistent with all unshaded regions, with dashed contours showing representative values (425 GeV, 450 GeV, 475 GeV). The outer shaded regions do not give rise to a two stage phase transition for the reasons outlined in the relativistic discussion. In the lower shaded region (green), the $X$ mass consistent with the asymmetry transfer is much larger than $m_\phi$, making the dark matter unstable.

We mark the regions where the mixing angle and thermal suppression terms in (\ref{eq:asymxnratio}) dominate. Dark matter masses  range from $400-500$ GeV with the parameters chosen. Comparing with Figure \ref{fig:Thermal}, the presence of the large mass mixing significantly changes the asymmetry, even in  regions where  thermal terms are largest. This favors heavy dark matter, $\mathcal O(400\,\,\rm GeV)$, as the large $X$ mass suppresses both the mixing angle and thermal contributions and results in the correct dark matter abundance, even in the presence of two sources of asymmetry transfer.

 The range of parameters with dominant mixing angle  is much more restricted than for relativistic or thermally-suppressed dark matter. The reason is that, except for small $k_2$, tunneling occurs shortly after the vacua become degenerate and when the $\phi$ VEV is small. This typically leads to a mixing angle that is smaller than the Boltzmann suppression factor.

\end{enumerate}

To summarize, with  a two Higgs model with a two stage phase transition, the relationship between the baryon and dark matter asymmetries can be determined by  thermal effects and/or the mixing angle. With large mass mixing ($y_X\gtrsim1$), the mixing angle contribution to the asymmetry is always important, and the ratio of $\Delta_X/\Delta_{\rm B}$ is modified from the predictions of earlier work where only thermal effects are relevant to the dark matter asymmetry \cite{Buckley:2010ui,Cohen:2009fz}. The model is consistent with a wide range of dark matter masses, from $3.3$ GeV in the relativistic case to 300-500 GeV in the thermally-dominated and mixing-angle-dominated regimes.

Note that for this model the transfer is always in equilibrium and the thermal scattering rate is always comparable to the oscillation rate.

\subsection{Annihilation of the symmetric component}\label{sec:annihilation}
In addition to generating the correct dark matter asymmetry, the coupling must be large enough that the symmetric component annihilates efficiently so that the late-time $X$ density is dominated by the asymmetric component. Our model provides a built-in mechanism for symmetric $X$ annihilation via the $t$-channel scalar exchange process process $X\bar X^{\dagger} \rightarrow L L^{\dagger}$. The cross section for this process is
\ben\label{eq:annxsec}
\sigma = \frac{y_X^4}{16\pi}\cdot\frac{m_X^2}{m_{\phi}^4}.
\een
The requirement that the symmetric number density of $X$ at freeze-out be less than the asymmetric component constrains the Yukawa coupling.  For masses in the $\mathcal O(100\,\,\mathrm{GeV})$ range, this requires $x\equiv m_X/T_{\rm freeze-out}\sim25$ and,
\ben
y_X^4 > (6.6\times10^3)\left(\frac{m_{\phi}^4\,e^x}{\MPl \, m_X^3\sqrt x}\right).
\een
For values $m_X\sim m_{\phi}\sim \mathcal O(500\,\,\mathrm{GeV})$, this gives the constraint $y_X > 0.45$. This is satisfied for the parameters considered above and is true over much of the parameter space consistent with asymmetric dark matter where the $X$ asymmetry comes from mass mixing.

When $m_X=3.3$ GeV, the suppression in the cross section from $(m_X/m_{\phi})^2$ can be very large. When $m_{\phi}$ is at its lowest bound of 90 GeV, then $y_X\gtrsim0.6$. For larger values of $m_{\phi}$, then $y_X>1$ to ensure the annihilation of the symmetric component. There may also exist other annihilation channels (for example, if there is a $\mathrm U(1)'$ under which $X$ and $\phi$ are charged) to further deplete the symmetric component.

\subsection{Phenomenology}\label{sec:2higgspheno}
While the dark matter in this model is a singlet under the Standard Model gauge group, the model also contains a new charged field, the doublet scalar $\phi$. This leads to a loop-suppressed coupling of $X$ to nucleons, as described in \cite{Raby:1987nb,Kaplan:2009ag}. The direct detection cross section is the same as \cite{Cohen:2009fz},
\ben
\sigma_{\rm dd}\approx (4\times10^{-46}\,\,\mathrm{cm}^2)\left(\frac{Z/A}{0.4}\right)^2\left(\frac{500\,\,\mathrm{GeV}}{m_{\phi}/y_X}\right),
\een
where $m_\phi/y_X\approx10-1000$ GeV in the two Higgs model ($m_\phi\approx100-500$ GeV and $y_X\approx0.5-10$). This cross section should be tested by the next generation of detectors. The one-loop interaction with nucleons is dominant over direct interactions with electrons in the detector, due to the small mass of the electron and therefore the small recoil energy \cite{Kopp:2009et}.

 The doublets can be pair produced at the LHC by $s$-channel weak gauge bosons. The neutral component $\phi^0$ decays to $X+\nu$, while the charged components decays to $X+\ell^{\pm}$. The decay of $\phi^{\pm}\rightarrow\phi^0+W^{\pm}$ is suppressed by the small mass splitting (induced by electroweak symmetry breaking loops) between components of the doublet.

The charged component is easiest to find. Given the approximate lepton-$X$ flavor symmetry, the signature is two oppositely-charged leptons of the same flavor plus missing energy. This is exactly analogous to searches for charged left-handed sleptons in the MSSM, with $X$ being analogous to the neutralino LSP. In the case of sleptons, assuming no degeneracy with the SLP ($m_{\chi^0}\ll m_{\tilde\ell}$), the LHC will be capable of finding left-handed sleptons at the LHC at the $5\sigma$ level with masses up to 350 GeV with an integrated luminosity of $100\,\,\mathrm{fb}^{-1}$ at 14 TeV \cite{Bityukov:1997ck}. In our case, the doublet masses are in the range $100-500$ GeV, so there is the possibility of detection at the LHC, although discovery will require a long running time.

This model can be differentiated from supersymmetric models because the decay chains should be different. Furthermore, this is a more challenging search than gluino or squark searches, so by the time of detection of the doublets, we should already know whether or not supersymmetry is present at the weak scale.

Other possible signatures are the production of two $\phi^0$, which results in a monojet + missing energy signature, or the production of $\phi^+\phi^0$, with a signature of one lepton + missing energy. Both are difficult to distinguish from the Standard Model background, and we find it unlikely that either is promising.

Our model is an asymmetric leptophilic dark matter model \cite{Cohen:2009fz} and can therefore have suppressed indirect detection signals. The fact that the late-time dark matter density is dominated by the asymmetric component means that there may be insufficient anti-dark-matter in the late universe for there to be an appreciable annihilation signal. As shown in earlier works, the symmetric component could, however, be replenished by a small $X$ Majorana mass or there could be some remnant of the symmetric component due to annihilation freeze-out \cite{Graesser:2011wi}. Therefore, the detection of excesses in cosmic spectra does not rule out this model. As shown in \cite{Cohen:2009fz}, the annihilation cross section (\ref{eq:annxsec}) for weak scale dark matter is $\langle\sigma_{\rm ann}v\rangle\approx10^{-24}\,\,\mathrm{cm}^3/$s, which is consistent with the PAMELA and Fermi anomalies for $m_X=400$ GeV, $m_{\phi}=500$ GeV and $y_X=1$ \cite{Cohen:2009fz}.

\section{ Mass Mixing via Planck-Scale-Suppressed Operators} \label{sec:planck}
\subsection{Overview}
Our second and more generic scenario exploits large VEVs that we expect in the early universe. Examples of fields that are likely to have large, non-zero values include  moduli, flat directions in supersymmetric theories, and Goldstone bosons. At finite temperature and density, the kinetic energies of relativistic fields also have sizable background values. Generically, all such fields should couple to any other existing fields through Planck-suppressed operators. Since we do not expect such operators to respect any particular global symmetry, these operators should lead to the mixing of  dark matter and lepton quantum  numbers. While theories with  moduli and flat directions are most likely supersymmetric, mixing due to background energy  is also  present in non-supersymmetric theories.

The key point to take away for all models of mixing by Planck-suppressed operators is that mixing between different sectors \emph{does} occur due to Planck-scale physics (assuming gauge-invariant combinations of fields). Therefore, asymmetry transfer between sectors also occurs if the asymmetry generation happens early in the universe's history. This is  the dominant effect if the dark and Standard Model sectors are very isolated, and is still expected to be present as a sub-leading effect if other fields mediate dark-matter-number-violating interactions between the two sectors.

We consider both fermionic and scalar dark matter candidates. Since lower-dimensional operators lead to less-suppressed mixing, scalar dark matter proves to be particularly important when the mixing would otherwise be too small, such as when induced by thermal kinetic terms and field VEVs, which are temperature-suppressed and therefore generically smaller than those for moduli. For operators involving string moduli and other flat directions, both fermion and scalar mixing can lead to a large transfer of asymmetry between sectors.

 For simplicity, we consider models in which the dark matter is a Standard Model gauge singlet in order to avoid direct detection bounds. This implies that the field mixing with the dark matter is also a Standard Model gauge singlet\footnote{The dark matter can carry a gauge charge as long as the field with which it mixes is also charged.}. One option for the field mixing with dark matter is the right-handed neutrino in a model with Dirac neutrino masses, which carries lepton number. While we choose this example to work out the concrete details of mass mixing, the general framework is also compatible with Majorana leptogenesis (or GUT baryogenesis) in theories with additional lepton-number-carrying gauge singlets. As will be discussed later, models with heavy moduli can dilute the dark matter and baryon asymmetries. In the framework of Dirac leptogenesis, the dilution can be compensated by resonance enhancement, which is not generic. A more general alternative is to have Affleck-Dine leptogenesis (or baryogenesis) along a flat-direction as a composition of singlet $L (B)$ and SM leptons (baryons). Alternatively, the moduli can be light enough to be stable over the lifetime of the universe.

We focus on the case where a $B-L$ symmetry is created via Dirac leptogenesis \cite{Dick:1999je}. In such models, no net lepton number is created but an equal and opposite lepton number is sequestered in the LH and RH neutrino sectors. The processes that equilibriate the two sectors have rates that are suppressed by neutrino masses and so do not come into equilibrium until well after the time of the electroweak phase transition. The asymmetry stored in the RH (s)neutrinos can then be transferred to the dark sector via mass mixing terms and provide for a connection between the dark matter and baryon asymmetries.

For our models, we take the field content to be the MSSM as well as the new components outlined below. First, there are the dark matter and singlet lepton fields:
\begin{itemize}

\item Three generations of singlet RH neutrino chiral superfields $N_i$.

\item Chiral superfields $X$ and $\bar X$, which comprise the dark matter.

\end{itemize}
We also assume in this model the existence of fields responsible for Dirac leptogenesis, as outlined in Appendix \ref{app:Diraclepto}:
\begin{itemize}

\item At least two generations of heavy vector-like doublets $\psi$ and $\bar\psi$, whose decay generates the $B-L$ asymmetry.

\item A singlet field $\chi$ which is necessary for generating an asymmetry in both LH and RH neutrinos.

\item A $\mathrm U(1)_N$ symmetry that forbids direct coupling between the LH Standard Model lepton $L$ and the RH neutrino $N$. It can also permit the annihilation of the symmetric component of $X$ if gauged. The gauge symmetry is anomaly-free with charges $Q(X)=1$, $Q(N)=-1$, $Q(\psi) = 1$ and $Q(\chi)=1$ with and two additional fields, either in the lepton or dark sectors.

\end{itemize}
Note that the new matter content is similar to that for models of Majorana leptogenesis (and a dark matter sector). But in this case the right-handed neutrinos are light and additional  fields are present to generate the $B-L$ asymmetry through their decay.

The tree-level, renormalizable superpotential is
\ben\label{eq:Wrenorm}
W = M_{\psi}\bar\psi\psi+ y_u Q H_u \bar u+y_d QH_d \bar d + y_N NH_u\psi+y_L L\chi\bar\psi.
\een
The model possesses a Peccei-Quinn-type symmetry that forbids $\bar XX$ mass terms. These mass scales can be generated dynamically at later times, possibly through SUSY breaking. We then expect that the  $X$ masses (for both the fermions and scalars) are ultimately set by the soft SUSY scale.

\subsection{Mass mixing formalism}\label{sec:massmixing}
In Section \ref{sec:twost}, we  worked in the mass eigenstate basis. This is appropriate because the mass mixing in the two Higgs case is nearly temperature-independent around the decoupling temperatures (particularly when compared to the rate of the first-order phase transition), making the mass eigenbasis temperature independent as well. The system also consists of a mixture of relativistic and non-relativistic states, and the mass basis adequately describes the Boltzmann suppression of the heavy states. Finally, the system is fully mixed, meaning that the rate of the asymmetry transfer given by (\ref{eq:transferrate}) is in equilibrium ($\gg H$) and the chemical potentials for $X$ and $N$ track their equilibrium values.

By contrast, the VEVs inducing mass mixing for these models with Planck-suppressed operators are  highly temperature-dependent. In fact, there is no phase transition, with VEVs typically proportional to some power of $T$ and turning off gradually.  Because the mass mixing changes with time, the mass basis is only an instantaneous eigenbasis. To avoid continually changing basis over the course of the time evolution, it is simplest to use interaction eigenstates.

At  early times when the Planck-suppressed mass mixing operators are active, the fields are also highly relativistic. This is precisely the limit in which the standard neutrino oscillation formalism is valid. This section is dedicated to the overview of how distributions of particles with mass mixing evolve over time. A good review of this material can be found in  \cite{Gelmini:1994az}. We first present fermion mixing and then discuss scalars.

We first review the story with pure coherent oscillation, which is directly analogous to neutrino oscillation. For Weyl fermions $\psi_{\alpha}$, $\alpha=1,\ldots,N$, the mass part of the Hamiltonian is
\ben
\mathcal H_{\mathrm{mass}} = \frac{1}{2}M_{\alpha\beta}\psi_{\alpha}\psi_{\beta} + \mathrm{h.c.}.
\een
     We assume no mass degeneracies in order to simplify the results. We use Greek letters to label flavor eigenstates and Latin letters to label mass eigenstates. The mass eigenvalues $M_i^2$ are found by diagonalizing the matrix $M^{\dagger}M$. The rotation matrix $U$ that diagonalizes $M$ satisfies
\ben
\sum_{\alpha, \beta}\,U^{\dagger}_{i\alpha}(M^{\dagger}M)_{\alpha\beta}U_{\beta j} = M_i^2\delta_{ij}.
\een
The relationship between the mass basis and the flavor basis is
\ben
|\psi_{\alpha}\rangle = \sum_i\, U_{\alpha i}|\psi_i\rangle.
\een

 During a particle's propagation between interaction points, it can be described by a wavefunction in standard quantum mechanics. After a time $t$, the probability of oscillation from flavor state $\alpha$ to $\beta$ is given by $\mathcal P_{\alpha\rightarrow\beta}(t) = |\langle \psi_{\beta}(t)|\psi_{\alpha}(0)\rangle|^2$. Expanding the states in the mass basis and performing space-time evolution using the operator $e^{-iHt+i\mathbf p\cdot\mathbf x}$ gives, in the relativistic limit,
\ben
\mathcal P_{\alpha\rightarrow\beta}(t) = \mathrm{Re}\left[ \sum_{i,j} \,U_{\alpha i}U_{\alpha j}^*U_{\beta i}^*U_{\beta j}\exp\left(-i\frac{M_i^2-M_j^2}{2E}t\right)\right],
\een
where $\alpha\neq\beta$. The case where $\alpha=\beta$ is simply found by subtracting the probability of oscillating into anything else.

For the special case of two fields, we can write $U$ in terms of a single mixing angle,
\ben
U = \left(\begin{array}{cc}\cos\theta & \sin\theta \\ -\sin\theta & \cos\theta \end{array}\right).
\een
The oscillation probability then reduces to the familiar form
\ben
\mathcal P_{\alpha\rightarrow\beta}(t) = \sin^2 2\theta\,\sin^2\left(\frac{M_2^2-M_1^2}{4E}t\right).\label{eq:nuosci}
\een

We compute the exact rate at which the asymmetry is transferred between sectors by mass mixing using the density matrix formalism for fermions and scalars. Before doing so, we present a good, approximate formula for the rate of asymmetry transfer that agrees with the density matrix results in two important limits discussed in Sections \ref{sec:densityscalars} and \ref{sec:densityfermions}. This formula will be useful in understanding how the asymmetry rate depends on the mass eigenvalues and mixing angle. We derived all of our numerical results in Section \ref{sec:planck} using the density matrix formalism.

In an environment where frequent scatterings occur, as is typical in the early universe, oscillation is interrupted by interactions with other states. This decoheres the system and can be thought of as collapsing the $\psi$ wavefunction into one of its interaction eigenstates. Interactions are occurring at a total average rate $\Gamma_0$,
\ben\label{eq:thermal}
\Gamma_0 = \frac{1}{2}\left(\Gamma_{\alpha}+\Gamma_{\beta}\right),
\een
where $\Gamma_{\alpha}$ is the total rate with which species $\alpha$ interacts with the thermal background. The mean free path is $1/\Gamma_0$. The rate that species $\alpha$ changes into species $\beta$ is going to be given by this interaction rate, multiplied by the probability of oscillation:
\ben\label{eq:asymtransfer}
\Gamma_{\alpha\rightarrow\beta} = \Gamma_0\,\sin^22\theta\,\sin^2\left(\frac{M_2^2-M_1^2}{4\,E\,\Gamma_0}\right) \equiv  \Gamma_0\,\sin^22\theta\,\sin^2\left(\frac{\epsilon}{2\Gamma_0}\right),
\een
where $\epsilon = (M_2^2-M_1^2)/2E$.  Equation (\ref{eq:asymtransfer}) is easy to understand and is a good approximation to the actual evolution of various species in a thermal background and we use it frequently. If $\Gamma_{\alpha\rightarrow\beta}>H$, the mixing is in equilibrium. Otherwise, we can obtain an order-of-magnitude estimate of the amount of transfer from species $\alpha$ to $\beta$ by the quantity $\Gamma_{\alpha\rightarrow\beta}/H$ evaluated at the time the asymmetry is generated in one sector.

The $\alpha\rightarrow\beta$ transfer can proceed in two different ways associated with different limits of (\ref{eq:asymtransfer}). Which regime the system is in is independent of whether or not the mixing is in equilibrium. When $\epsilon/\Gamma_0\gg1$, the oscillation length is much shorter than the mean free path and so a larger amount of $\alpha$ particles are transferred per collision. As a result, fewer collisions are necessary to bring the system into equilibrium. The opposite limit, $\epsilon/\Gamma_0\ll1$, occurs when collisions occur very rapidly compared to the oscillation time scale and few $\alpha$ particles are converted to $\beta$ particles per collision. Nevertheless, the transfer process can still be in equilibrium if the mixing angle is large. We encounter examples of both limiting cases in Sections \ref{sec:moduli} and \ref{sec:bkd}.

The evolution of the different flavor states, taking into account oscillations and coherence-destroying thermal interactions, can be performed in a unified framework using the density matrix formalism \cite{Manohar:1986gj,Foot:1995qk,Dolgov:2003sg}. The system begins in some state $\rho_0$, where the diagonal components give the iniital relative abundances of each particle type and the off-diagonal components represent any coherences present in the initial state.
The time evolution of the density matrix is given by the equation
\ben\label{eq:densityevolve}
i\dot\rho = [H^{(\mathrm R)},\rho] -i\{H^{(\mathrm I)},\rho\},
\een
where $H^{(\mathrm R)}$ is the real part of the Hamiltonian calculated to leading order and includes masses and interactions with the thermal background, while $H^{(\mathrm I)}$ consists of the imaginary part of the background potential consisting of absorptive interactions with the background that cause the system to decohere. Integrating (\ref{eq:densityevolve})  yields the final abundance of each species. In Sections \ref{sec:densityscalars} and \ref{sec:densityfermions}, we give (\ref{eq:densityevolve}) in component form for scalar and fermion mixing and show that, in the limit of time-independent temperature and mixing, the result reduces to (\ref{eq:asymtransfer}).

We are interested in determining the evolution of an asymmetry that is stored entirely in one sector and subsequently migrates to a different sector through oscillations. Since the evolution equations are linear, it is appropriate to consider only the asymmetric components. The initial density matrix at the time the asymmetry is generated and that we use in our analysis is $\rho_0 = \mathrm{diag}(1,0,0,\ldots,0)$, where the 1 is the diagonal component of the species housing the asymmetry.

At finite temperature and density, particularly temperatures where $T>M_i$, finite temperature contributions to the background potential must be included in the mass mixing formulae (we assume that both sectors are in kinetic equilibrium so that there is a common temperature between them). Considering again the $2\times2$ case, the effective masses from thermal corrections for both fermions and scalars in the interaction basis are
\ben
|M(T)|^2-|M(0)|^2 = \left(\begin{array}{cc} \lambda_1^2T^2 & 0 \\ 0 & \lambda_2^2T^2 \end{array}\right),
\een
where $\lambda_1$ and $\lambda_2$ are effective couplings parameterizing all interactions with the background plasma\footnote{For fermions, finite temperature corrections do not give rise to a mass term in the Lagrangian due to chiral symmetry, but they do give rise to an effective potential that can be considered as a contribution to the $M^{\dagger}M$ matrix}. In Section \ref{sec:moduli}, we relate $\lambda_1$ and $\lambda_2$ to the couplings of specific models.

Using $E\approx 3T$ at finite temperature and density in the ultra-relativistic limit, we get the Hamiltonian
\ben\label{eq:hamiltonian}
H_{\alpha\beta} = 3T\,\delta_{\alpha\beta} + \frac{|M(T)|_{\alpha\beta}^2}{6T}.
\een
In the case that there exist common interactions between the sectors that are in thermal equilibrium, $T$ is the same for all fields. The mixing angle and energy eigenvalue splitting are unchanged when a term proportional to the identity is subtracted, so we can subtract the $T\,\delta_{\alpha\beta}$ term from (\ref{eq:hamiltonian}).

We now derive results for the particular cases of scalars and fermions.

\subsubsection{Scalars}\label{sec:densityscalars}

Let the tree-level, diagonal elements of $M_{ij}$ be denoted by $\mu_i$. Then, the total mass-squared matrix  is
\ben
|M(T)|^2 = \left(\begin{array}{cc} \mu_1^2+\lambda_1^2 T^2 & M_{12}^2 \\ M_{12}^2 & \mu_2^2 +\lambda_2^2T^2\end{array}\right).\label{eq:scalarmix}
\een
 The mass eigenvalues and mixing angle are
\bean
2m_{\pm}^2 &=& \lambda_1^2T^2+\lambda_2^2T^2+\mu_1^2+\mu_2^2\pm\sqrt{(\lambda_{21}^2T^2+\mu_2^2-\mu_1^2)^2+4M_{12}^4},\label{eq:sceps}\\
\sin\theta &=& \frac{\sqrt 2M_{12}^2}{\sqrt{(\lambda_{21}^2T^2+\mu_2^2-\mu_1^2)(\lambda_{21}^2T^2+\mu_2^2-\mu_1^2+\sqrt{(\lambda_{21}^2T^2+\mu_2^2-\mu_1^2)^2+4M_{12}^4})+4M_{12}^4}}\label{eq:scmixing},
\eean
where $\lambda_{21}^2=\lambda_2^2-\lambda_1^2$.

The Hamiltonian (\ref{eq:hamiltonian}) is
\ben
H = \frac{1}{6} \left(\begin{array}{cc} \mu_1^2+\lambda_1^2 T^2 &M_{12}^2 \\ M_{12}^2 & \mu_2^2 +\lambda_2^2T^2\end{array}\right).
\een
and
\ben
\rho = \left(\begin{array}{cc} \rho_{11} & \rho_{12} \\  \rho_{21} & \rho_{22}\end{array}\right).
\een
The density matrix evolution equations (\ref{eq:densityevolve}) in component form are \cite{Dolgov:2003sg}
\ben\label{eq:rhoscalar}
\left(\begin{array}{c} \dot\rho_{11}\\\dot\rho_{22}\\\dot\rho_{12}\\\dot\rho_{21}\end{array}\right) = \frac{1}{6T}\left(\begin{array}{cccc} 0 & 0 & iM_{12}^2 & -iM_{12}^2 \\
0& 0 & -iM_{12}^2 & iM_{12}^2 \\
iM_{12}^2 & -iM_{12}^2 & -2\Gamma_0T+i\left[\mu_2^2-\mu_1^2-\lambda_{21}^2T^2\right] & 0 \\
-iM_{12}^2 & iM_{12}^2 & 0 & -2\Gamma_0T -i\left[\mu_2^2-\mu_1^2-\lambda_{21}^2T^2\right]
\end{array}\right)\left(\begin{array}{c} \rho_{11}\\\rho_{22}\\\rho_{12}\\\rho_{21}\end{array}\right).
\een
In this case, (\ref{eq:thermal}) gives
\ben
\Gamma_0 = \frac{1}{2}\left(\Gamma_1+\Gamma_2\right).
\een
Note that the previous formula applies also in the limit that  $M_{12}$ and $T$ are independent of time, in which case we have a linear system of first-order differential equations. The solutions are   exponential functions, with the coefficients in the exponential being given by the eigenvalues of the matrix in (\ref{eq:rhoscalar}). There is one zero eigenvalue, representing the equilibrium distribution, and the approach to chemical equilibrium (i.e.~the asymmetries being equal in both fields) is given by the smallest of the eigenvalues, denoted by $\gamma_{\mathrm{slow}}$. Defining
\ben
\epsilon\equiv \frac{m_+^2-m_-^2}{6T},
\een
the solutions can be found analytically in the limits $\Gamma_0\ll \epsilon$ and $\Gamma_0\gg\epsilon$. The solutions are
\bean
\gamma_{\mathrm{slow}} &=& 2\Gamma_0\sin^2\theta\cos^2\theta ,\,\,\,\,\,\,\,\,\Gamma_0\ll\epsilon,\\
\gamma_{\mathrm{slow}} &=& \sin^2\theta\cos^2\theta\,\frac{\epsilon^2}{\Gamma_0},\,\,\,\,\,\,\,\,\,\Gamma_0\gg\epsilon,
\eean
which reproduces (\ref{eq:asymtransfer}) in the same limits. This confirms our physical picture of transfer via mass mixing and interactions with the background, and also checks  equation (\ref{eq:asymtransfer}).

Since all of the parameters in our theory are actually time-dependent, the system never reaches chemical equilibrium if $\gamma_{\rm slow}<H$. There is a significant parameter space for the models in Sections \ref{sec:moduli} and \ref{sec:bkd} for which this is true. In this case, the amount of asymmetry transferred between sectors is approximately $\gamma_{\rm slow}/H$.

\subsubsection{Fermions}\label{sec:densityfermions}
Suppose we have three Weyl fermions with a Dirac mass $\mu$ between $\psi_1$ and $\psi_2$ and a Dirac mass mixing $M_{13}$ between $\psi_1$ and $\psi_3$. The mass-squared matrix is
\ben
|M(T)|^2 = \left(\begin{array}{ccc} \mu^2+M_{13}^2+\lambda_1^2T^2 & 0 & 0 \\
0 & \mu^2+\lambda_2^2T^2 & \mu M_{13} \\
0 & \mu M_{13} & M_{13}^2 + \lambda_3^2T^2\end{array}\right).\label{eq:fermimix}
\een
Unlike in the scalar case, the tree-level mass $\mu$ appears in the off-diagonal terms. The matrix is block diagonal and only the lower $2\times2$ sector is mixed. The mass eigenvalues and mixing angle are
\bean
2m_{\pm}^2 &=&\mu^2+M_{13}^2+(\lambda_2^2+\lambda_3^2)T^2\pm\sqrt{\mu^4+\left(M_{13}^2+\lambda_{32}^2T^2\right)^2+2\mu^2(M_{13}^2-\lambda_{32}^2T^2)},\label{eq:fermieps}\\
\sin\theta &=& \frac{\sqrt 2 \mu M_{13}}{\sqrt{A+M_{13}^4+(\mu^2-\lambda_{32}^2T^2-M_{13}^2)\sqrt{\mu^4+2\mu^2(M_{13}^2-\lambda_{32}^2T^2)+(M_{13}^2+\lambda_{32}^2T^2)^2}}},\label{eq:fermimixing}
\eean
where
\ben
A = (\mu^2-\lambda_{32}^2T^2)^2+2M_{13}^2(\mu^2+\lambda_{32}^2T^2).
\een
The Hamiltonian is
\ben
H = \frac{1}{6T}\left(\begin{array}{ccc} \mu^2+M_{13}^2+\lambda_1^2T^2 & 0 & 0 \\
0 & \mu^2+\lambda_2^2T^2 & \mu M_{13} \\
0 & \mu M_{13} & M_{13}^2 + \lambda_3^2T^2\end{array}\right).
\een
 The density matrix is
 \ben
 \rho = \left(\begin{array}{ccc} \rho_{11} & \rho_{12} & \rho_{13}\\
 \rho_{21} & \rho_{22} & \rho_{23}\\
 \rho_{31} & \rho_{32} & \rho_{33}\end{array}\right).
 \een
Because $H$ is block diagonal, the density matrix evolution equations involving $\psi_1$ decouple from those independent of $\psi_1$, giving three independent systems of equations,
\bean\label{eq:rhofermion}
\left(\begin{array}{c} \dot\rho_{22}\\\dot\rho_{33}\\\dot\rho_{23}\\\dot\rho_{32}\end{array}\right) &=& \frac{1}{6T}\left(\begin{array}{cccc} 0 & 0 & i\mu M_{13} & -i\mu M_{13} \\
0& 0 & -i\mu M_{13} & i\mu M_{13} \\
i\mu M_{13} & -i\mu M_{13} & -6\Gamma_0T+i(\mu^2-M_{13}^2-\lambda_{32}^2T^2) & 0 \\
-i\mu M_{13} & i\mu M_{13} & 0 & -6\Gamma_0T -i(\mu^2-M_{13}^2-\lambda_{32}^2T^2)
\end{array}\right)\left(\begin{array}{c} \rho_{22}\\\rho_{33}\\\rho_{23}\\\rho_{32}\end{array}\right),\nonumber\\
\left(\begin{array}{c} \dot\rho_{12}\\\dot\rho_{13}\end{array}\right) &=& \frac{1}{6T}\left(\begin{array}{cc}-6\Gamma_0'T -iM_{13}^2 & i\mu M_{13}  \\
i\mu M_{13}&-6\Gamma_0''T -i\mu^2 \end{array}\right)\left(\begin{array}{c} \rho_{12}\\\rho_{13}\end{array}\right),\nonumber\\
\dot\rho_{11}&=&0.\label{eq:rhoevolve}
\eean
The equations for $\rho_{21}$ and $\rho_{31}$ are found by taking the conjugate of those for $\rho_{12}$ and $\rho_{13}$. Because there are now three particle types, we have three rates and
\bean
\Gamma_0 &=& \frac{1}{2}\left(\Gamma_2+\Gamma_3\right),\\
\Gamma'_0 &=& \frac{1}{2}\left(\Gamma_1+\Gamma_2\right),\\
\Gamma''_0 &=& \frac{1}{2}\left(\Gamma_1+\Gamma_3\right).
\eean
Since $\rho_{11}$ is conserved, we can only have transfer between particles of species 2 and 3, and therefore only the first system of equations in (\ref{eq:rhoevolve}) is relevant for our purposes.

 As in the scalar case, the analytic solution to (\ref{eq:rhoevolve}) in the case of time-independent $T$ and $M_{13}$ is
 \bean
\gamma_{\mathrm{slow}} &=& 2\Gamma_0\sin^2\theta\cos^2\theta ,\,\,\,\,\,\,\,\,\Gamma_0\ll\epsilon,\\
\gamma_{\mathrm{slow}} &=& \sin^2\theta\cos^2\theta\,\frac{\epsilon^2}{\Gamma_0},\,\,\,\,\,\,\,\,\,\Gamma_0\gg\epsilon,
\eean
which reproduces (\ref{eq:asymtransfer}).

\subsection{Moduli-induced mass mixing}\label{sec:moduli}
\subsubsection{Overview}
One class of fields that are expected to have large background values at early times are moduli fields; examples include string moduli or Polonyi fields for supersymmetry (SUSY) breaking. As their potentials are exactly flat at the perturbative level in SUSY, moduli can be found at finite temperature far from their zero-temperature minima. Such fields can have VEVs as large as $\MPl$ following inflation. The moduli subsequently roll down to their true minima, which would be established by SUSY-breaking terms in the potential. During this period of rolling, the VEVs can induce large mass mixing terms between other fields in the theory.

Note that moduli decay can create entropy that would dilute both lepton and dark matter numbers. We assume that either decay occurs very late (as with light moduli) with Dirac leptogenesis or that a more efficient means of baryogenesis (such as Affleck-Dine baryogenesis) applies. Dirac leptogenesis can be efficient as well with resonance enhancement, but it is less generic.  We show that both light and heavy moduli are possible in Section \ref{sec:dilution}. For clarity, we use bold-face when discussing the cases of heavy and light moduli.

We consider the dark matter to be a vector-like pair of superfields $X,\bar{X}$ with tree-level mass $\mu_{\rm tree}$. Mass mixing arises from Planck-suppressed higher-dimensional operators.
Mass mixing between $X$ and $N$ generically occurs for both fermions and their scalar counterparts.

For {\bf heavy moduli}, mixing between fermion components and a modulus $\phi$,  mixing can come from superpotential terms such as
\ben\label{eq:fermionmixing}
\Delta W = \frac{c\,\phi^2}{\MPl}XN+\mathrm{h.c.},
\een
which gives a mixing
\ben\label{eq:fermionmassmixing}
m_{XN} = \frac{c\,\phi^2}{\MPl}.
\een
Since the Hamiltonian contains the squared mass matrices, the relevant mixing term is not $m_{XN}$ but the product
\ben\label{eq:fermionsquared}
m_{\rm mix}^2 = \mu_{\rm tree}\,m_{XN} = \frac{c\,\mu_{\rm tree}\phi^2}{\MPl}.
\een
The dark matter is stable if $\mu_{\rm tree}<2m_\phi$.

Superpotential terms also generate mixing between the scalar components of $X$ and $N$. In the presence of a spurion field $S$ with SUSY-breaking $\mathcal F$-term, an allowed operator is
\ben\label{eq:scalarmixing}
 \Delta W=\frac{c_{\rm s}\,S\phi^2}{\MPl^2}XN+\mathrm{h.c.},
\een
  If $\mathcal F_S$ is responsible for SUSY breaking in the MSSM, the mixing is
\ben\label{eq:scalarmassmixing}
m_{XN}^2 =\frac{c_{\rm s}\,m_{3/2}\,\phi^2}{\MPl}.
\een
We compare the mixing term in the squared mass matrix (\ref{eq:scalarmassmixing}) with the fermion case (\ref{eq:fermionsquared}). The ratio of fermion to scalar mixing is $\mu_{\rm tree}/m_{3/2}$.

The late-time dark matter particle can be either scalar or fermion, depending on the details of the model. It is, however, irrelevant at late times whether the dominant mixing was through scalars or fermions, since interactions within the $X$ sector  (most importantly, those responsible for annihilation of the symmetric component)  equilibriate the asymmetry among the scalar and fermion components.

We expect that higher-dimensional couplings to moduli fields  also contribute to $X\bar X$ mass terms. For the fermion case,
\ben
 \Delta W=\frac{c_X\,\phi^2}{\MPl}X\bar X+\mathrm{h.c.}.
\een
This operator doesn't violate dark matter or lepton numbers, but is relevant to the mixing angle. The resulting diagonal mass is
\ben
\mu_X = \mu_{\rm tree} + \frac{c_X\,\phi^2}{\MPl} \equiv \mu_{\rm tree} + \kappa\,m_{XN}.
\een
Since both $X\bar X$ and $XN$ get masses from the moduli VEV, we express everything in terms of $m_{XN}$ and the proportionality constant $\kappa = m_{X\bar X}/m_{XN}$, which we expect to be $\mathcal O(1)$ since both masses arise from the same moduli fields and operators of the same dimension. For scalars, the corresponding diagonal mass term is
\ben
\mu_{X\,\rm{scalar}}^2 = \mu_{\rm tree}^2+ \frac{\kappa\,c_{\rm s}\,m_{3/2}\,\phi^2}{\MPl}.
\een
In the scalar case, the diagonal mass does not contribute to the off-diagonal component of the mass-squared matrix, $m_{XN}^2$. We show in Section \ref{sec:approximations} that the asymmetry transfer rate does not depend on $\mu_{X\,\rm{scalar}}^2$,

For {\bf light moduli}, the condition $m_X<2m_\phi$ is not satisfied. The fermion decay rate is.
\ben
\Gamma_X \approx \frac{m_X^3}{\MPl^2},
\een
and the $X$ lifetime would be much below the bound of $\tau > 10^{26}$ seconds  \cite{Meade:2009iu}. For this model, we assume a $Z_3$ symmetry that only allows higher-dimensional operators such as
\ben\label{eq:fermionmixinglight}
\Delta W_{\rm light} = \frac{c\,\phi^3}{\MPl^2}XN+\mathrm{h.c.},
\een
which gives a mixing
\ben\label{eq:fermionmassmixinglight}
m_{XN} = \frac{c\,\phi^3}{\MPl^2}.
\een
The diagonal component of the mass-squared matrix is
\ben
m_{\rm mix}^2 = \frac{c\,\mu_X\phi^3}{\MPl^2}.
\een
The corresponding superpotential term for scalar mixing is
\ben
\Delta W = \frac{c\,S\phi^3}{\MPl^3}XN+\mathrm{h.c.},
\een
which gives a mixing term in the mass-squared matrix
\ben\label{eq:scalarmassmixinglight}
m_{XN}^2 = \frac{c\,m_{3/2}\phi^3}{\MPl^2}.
\een
As with heavy moduli, the ratio of fermion to scalar mixing is $\mu_X/m_{3/2}$. The decay width is
\ben
\Gamma_X \approx \frac{m_X^5}{\MPl^4},
\een
safely satisfying the constraints for $m_X\sim$ GeV-TeV. The mixing term for scalars is the SUSY-breaking spurion $S$ coupled to (\ref{eq:fermionmixinglight}).

We present a toy model for the potential of {\bf moduli of any mass} \cite{Dine:1995kz}:

\ben\label{eq:modpotential}
  V=(m_\phi^2-a^2H^2)|\phi|^2+\frac{1}{2\MPl^2}(m_\phi^2+b^2H^2)|\phi|^4,
\een
where $H$ is the Hubble scale. The $m_\phi^2$ and $H^2$ mass terms arise from SUSY breaking due to MSSM SUSY breaking terms and finite-temperature effects, respectively. At early times, the minimum of this potential lies at
\ben
  \langle\phi\rangle=\MPl\sqrt{\frac{a^2H(t)^2-m_\phi^2}{b^2H(t)^2+m_\phi^2}}.
\een
At the critical time $t_{\mathrm c}$, when $H(t_{\mathrm c}) = m_\phi/a$, the minimum of the potential disappears. Because of damping in the equations of motion, $\phi$ is unable to efficiently track the minimum of the potential near $t_{\mathrm c}$ and is left at some finite value. After the critical time, it begins to oscillate with a power-law-damped envelope, rolling toward the true minimum of $\phi=0$.
Its equation of motion is
\ben\label{eq:phieom}
\ddot{\phi}+3H\dot{\phi}+2(m_\phi^2-a^2H^2)\phi+\frac{2}{\MPl^2}(m_\phi^2+b^2H^2)\phi^3=0
\een

To obtain the asymmetry transfer rate, we must also specify the interactions in the theory. The simplest model that allows for the annihilation of the symmetric component of $X$ is a $\mathrm U(1)'$ gauge interaction with coupling $g$ under which $X$ and $N$ are oppositely charged. In the most minimal scenario, $N$ also has Yukawa interactions (with coupling $y$) from leptogenesis. We then have, in the notation of section \ref{sec:massmixing},
\bean\label{eq:couplings}
\lambda_X^2 &=& g^2/8,\\
\lambda_N^2 &=& g^2/8+y^2/8,\\
\lambda_{NX}^2 &=& y^2/8.
\eean

The Yukawa coupling $y$ depends on the model of leptogenesis. For Dirac leptogenesis, the effective coupling is $y=\mathrm{Max}(y_L^2,y_N^2)T_{\rm lep}^2/M_\psi^2$ with $T_{\rm lep}<M_\psi$. For Affleck-Dine leptogenesis, $y=(T_{\rm lep}/\MPl)^n$, where $n$ is the dimension of the operator lifting the flat direction.

Because of its importance, we repeat the analytic estimate of the transfer rate (\ref{eq:asymtransfer}):
\be
\Gamma_{N\rightarrow X} = \Gamma_0\sin^22\theta\sin^2\left(\frac{\epsilon}{2\Gamma_0}\right).
\ee
Three parameters  influence the transfer rate: the mass eigenstate energy splitting $\epsilon=(m_{+}^2-m_{-}^2)/6T$, the mixing angle $\theta$ (both of which are explicitly given in eqs.(\ref{eq:fermieps},\ref{eq:fermimixing})), and the thermal scattering rate $\Gamma_0\approx \frac{1}{8\pi^3}(g^4+y^2)T$. The asymmetry transfer rate is reduced for small mixing angles and small mass splittings. Note that in the limit where $\epsilon\gg\Gamma_0$, the oscillation is rapid and $\sin^2\left(\frac{\epsilon}{2\Gamma_0}\right)\rightarrow 1/2$ takes on its average value.

There are constraints associated with the couplings and scales in theory. They are:

\begin{itemize}

 \item Leptogenesis should occur after reheating to avoid diluting the lepton asymmetry with the massive quantity of entropy released by inflaton decay.

 \item In order to ensure efficient depletion of the symmetric component of $X$, the $X$ annihilation coupling should be $g\sim\mathcal O(1)$.

 \item $y\ll1$ for the heavy field in leptogenesis to decay out of equilibrium.

 \item $T>m_{XN}$ to avoid thermal suppression of $X$ and $N$ fields.

 \end{itemize}

\subsubsection{Evolution of $\phi$}\label{sec:phievolution}
The evolution of $\phi$ (\ref{eq:phieom}) depends on its mass. We present the evolution of $\phi$ for both heavy moduli ($m_\phi \gtrsim50$ TeV) and light moduli ($m_\phi\lesssim10$ MeV). We justify these mass ranges in Section \ref{sec:dilution}.

We first consider {\bf heavy moduli}, $\phi$ oscillation begins when $H\approx m_\phi$. For $m_\phi\gtrsim50$ TeV and $T_{\rm RH}\lesssim10^{12}$ GeV, oscillation will begin prior to inflaton decay.  In between the time $t_c$ when oscillation begins and reheating, the cosmic background density is matter dominated by inflaton and moduli fields that scale as massive matter. This gives an approximate solution to the $\phi$ evolution:
\ben\label{eq:phisol}
\phi\approx\phi_0\frac{\sin(m_\phi t)}{(m_\phi t)},
\een
where $\phi_0\sim \MPl$.

At the critical time $t_{\rm c}$, $H\approx m_\phi$ and $\langle\phi\rangle = \MPl$, so the $\phi$ and inflaton fields have comparable energy densities. Before the inflaton decays, both $\rho_\phi$ and $\rho_{\rm inf}$ scale identically with $T$, so their densities are also  comparable at the time of reheating. Setting $\rho_\phi=\rho_{\rm inf}$ at the time of reheating establishes the relationship between temperature and time in the subsequent $\phi$-dominated stage of the universe's evolution. Assuming the inflaton decays entirely to radiation, equating the inflaton and moduli densities gives
\ben
\frac{\pi^2}{30}g_*T_{\rm RH}^4 \approx m_\phi^2\,\phi(t_{\rm RH})^2.
\een
The background remains matter dominated and the relation (\ref{eq:phisol}) stays valid until the time of moduli decay, and during this epoch the temperature-time relation is
\ben\label{eq:Ttrelation}
T=T_{\rm RH}(t_{\rm RH}/t)^{2/3}.
\een

The solution to the evolution of $\phi$ for heavy moduli is
\ben\label{eq:dMmoduli}
\phi \approx \frac{a}{b}\,\MPl\left(\frac{t_c}{t}\right) \approx \frac{a}{b}\,\MPl\left(\frac{t_c}{t_{RH}}\right)\left(\frac{T}{T_{RH}}\right)^{3/2}.
\een
The mass mixing follows from (\ref{eq:fermionmassmixing}) and (\ref{eq:scalarmassmixing}).
Asymmetry transfer from $N$ to $X$ begins only at the time at which leptogenesis occurs, $t_{\rm lep}$. Because $m_{XN}$ turns off as $T^3$, the dark matter asymmetry is essentially established at $t_{\rm lep}$.
The mass mixing values $m_{XN}$ clearly depend on the model input parameters $a, b, c$, and $t_{\rm lep}$.

We now turn to {\bf light moduli}. In this case, oscillation must begin before $H=m_\phi$ to avoid having $\phi$ overclose the universe. One possibility is that the $\phi$ negative mass-squared predominantly comes from inflaton couplings that decay at reheating (see Section \ref{sec:dilution}) for details. Oscillation then begins at the time of reheating.

For all values of $T_{\rm RH}$ and light $m_\phi$ satisfying the overclosure bound (\ref{eq:overclosure}), oscillation begins only after reheating. As a result, the universe is radiation-dominated during the time of leptogenesis and mass-mixing, in contrast to the matter-dominated scenario for heavy moduli. In a radiation-dominated universe, $T$ and $t$ are related by \cite{Kolb:1990vq}
\ben
t = \frac{0.301\MPl}{\sqrt{g_*}\,T^2}
\een
and the $\phi$ evolution is
\ben
\phi \approx \frac{a}{b}\,\MPl\left(\frac{t_{\rm RH}}{t}\right)^{3/4}  \approx \frac{a}{b}\,\MPl\left(\frac{T}{T_{\rm RH}}\right)^{3/2}.
\een
The mass mixing is obtained from (\ref{eq:fermionmassmixinglight}) for fermions and (\ref{eq:scalarmassmixinglight}) for scalars.

\subsubsection{Analytic approximations}\label{sec:approximations}

We derive approximations that directly link the parameters from moduli mixing to the mixing angle and mass eigenvalue splittings in Section \ref{sec:massmixing}. There are three relevant limits for the mass eigenvalue splitting $\epsilon$ and $\sin\theta$. These are $m_{XN}\gg \mu_X$, $m_{XN}\ll\mu_X$, and $m_{XN}\sim\mu_X$.

The results for fermions follow from equations (\ref{eq:fermieps}) and (\ref{eq:fermimixing}):
\bean
\epsilon = \frac{\mu_X^2}{6T},&\,\,\,\sin\theta = \frac{m_{XN}}{\mu_X},&\,\,\,\mu_X\gg m_{XN};\\
 \epsilon = \frac{m_{XN}^2}{6T},&\,\,\,\sin\theta = \frac{\mu_X}{m_{XN}},&\,\,\,\mu_X\ll m_{XN};\\
\epsilon = \frac{m_{XN}^2}{3T},&\,\,\,\sin\theta = \frac{1}{\sqrt 2},&\,\,\,\mu_X= m_{XN}.
\eean
The overall thermal interaction rate is $\Gamma_0 \approx g^4 T/8\pi^3$. When the gauge coupling is $\mathcal O(1)$ (as needed for annihilation of the symmetric component of dark matter),  $m_{XN}$, $\mu_X<T$ in order to avoid thermal suppression. This means that typically $\epsilon \lesssim \Gamma_0$ and the asymmetry transfer rate in each limit (\ref{eq:asymtransfer}) reduces to the same quantity,
\ben
\Gamma_{N\rightarrow X} = \frac{4\sin^2\theta\cos^2\theta\epsilon^2}{2\Gamma_0} \approx \frac{m_{XN}^2\,\mu_X^2}{18T^2\,\Gamma_0}\approx \frac{\mu_X^2\phi^4}{3T^2\,\Gamma_0\MPl^2}
\een
For $\kappa=1$ (i.e.~including higher-dimensional corrections to the diagonal mass) and $m_{XN}\gg \mu_{\rm tree}$, this rate $>H$ and the system is in chemical equilibrium for $m_{XN}\sim T$,  while it  is out of chemical equilibrium for lower mixings. In this limit,
\ben
\Gamma_{N\rightarrow X} \approx \frac{\phi^8}{3T^2\,\Gamma_0\MPl^4}.
\een
Since $\phi$ is proportional to $(T/T_{\rm RH})^3$, if reheating and leptogenesis occur around the same scale, the asymmetry transfer is in equilibrium and chemical equilibrium is reached. If, however, there is a large hierarchy between $T_{\rm RH}$ and $T_{\rm lep}$, the transfer rate is suppressed.

In the case of scalars, from equations (\ref{eq:sceps}) and (\ref{eq:scmixing}):
\bean
\epsilon = \frac{\mu_X^2}{6T},&\,\,\,\sin\theta = \frac{m_{XN}^2}{2\mu_X^2},&\,\,\,\mu_X\gg m_{XN};\\
 \epsilon = \frac{m_{XN}^2}{3T},&\,\,\,\sin\theta = \frac{1}{\sqrt 2},&\,\,\,\mu_X\ll m_{XN};\\
\epsilon = \frac{\sqrt 5m_{XN}^2}{6T},&\,\,\,\sin\theta = \frac{\sqrt 2}{\sqrt{5-\sqrt 5}},&\,\,\,\mu_X= m_{XN}.
\eean
For similar reasons to the fermion case, when $m_{XN}$, $\mu_X<T$, the asymmetry transfer rate (\ref{eq:asymtransfer}) in all three limits is approximately
\ben
\Gamma_{N\rightarrow X} = \frac{4\sin^2\theta\cos^2\theta\epsilon^2}{2\Gamma_0} \approx \frac{m_{XN}^4}{18T^2\,\Gamma_0}\approx \frac{m_{3/2}^2\phi^4}{18T^2\,\Gamma_0\MPl^2}.
\een
Once again, when leptogenesis occurs close to the scale of reheating, $m_{XN}$ is large and the system is in chemical equilibrium. At lower leptogenesis scales and smaller values of $m_{XN}$, the system is out of chemical equilibrium.

We can compare the transfer rate from mixing between fermions and scalars:
\ben\label{eq:ratecompare}
\frac{\Gamma_{N\rightarrow X,\,\mathrm{scalar}}}{\Gamma_{N\rightarrow X,\,\mathrm{fermion}}} = \frac{m_{3/2}^2}{\mu_X^2}.
\een
Therefore, scalars  give the dominant contribution to mixing if $m_{3/2}\gg \mu_X$, and fermions give the dominant contribution if $\mu_X\gg m_{3/2}$. Of course, the heavier of the two state will decay to the lighter state, which is the dark matter.

\subsubsection{Numerical results}

To confirm our approximate results, we numerically solve the density matrix evolution equations to determine the magnitude of the $N\rightarrow X$ asymmetry transfer. We summarize the results below. The dark matter density is $\Omega_{\rm DM} \approx(5-6)\Omega_{\rm B}$. We choose parameters so that $\Omega_{\rm DM} = 5\Omega_{\rm B}$.

We begin with {\bf heavy moduli}. $\phi$ oscillation begins before reheating for $m_\phi \gtrsim$ 50 TeV, and $\phi$ is damped from its initial value of $\MPl$ by the time of reheating. For $T_{\rm RH}\sim10^{10}$ GeV, reheating occurs only shortly after oscillation begins and $\phi$ is not much damped from $\MPl$. As a result, fermion mass mixing is $>T$ and fermionic modes are suppressed, allowing the scalar mixing to dominate the asymmetry transfer.

With smaller reheating temperatures, $T_{\rm}\sim10^8$ TeV, $\phi$ is sufficiently damped at reheating that fermionic modes are not thermally suppressed. In this case, either scalar or fermion mixing can dominate, depending on the ratio of $\mu_X/m_{3/2}$, according to (\ref{eq:ratecompare}).

We present our results in terms of benchmark points for a high reheat temperature of $T_{\rm RH}=10^{10}$ GeV and a lower reheat temperature of $T_{\rm RH}=10^8$ GeV. We use $m_\phi=50$ TeV for both cases.

For $T_{\rm RH}=10^8$ GeV and all leptogenesis scales $T_{\rm lep}<T_{\rm RH}$, fermion mixing $m_{XN}<T$ and so we must consider both scalar and fermion mixing. With $a=b=1$, $c=\kappa=2$, some benchmark points are
\begin{enumerate}

\item $T_{\rm lep}=10^8$ GeV and $m_X=5$ GeV. The dominant mixing is from fermions because\\
 $\mu_X \approx5\times10^5\,\,\mathrm{GeV}> m_{3/2}\approx 5\times10^4$ GeV.
\item $T_{\rm lep} = 10^7$ GeV and $m_X=90$ GeV. The dominant mixing is from scalars because\\
 $\mu_X\approx 90$ GeV $\ll m_{3/2}$.

\end{enumerate}
The transfer rate is fast enough to bring $X$ and $N$ into chemical equilibrium for point \#1 above, consistent with $\sim5$ GeV dark matter. The transferred asymmetry is suppressed for point \#2, consistent with heavier dark matter. Heavier masses $\gtrsim100$ GeV are consistent with later leptogenesis times.

For $T_{\rm RH}=10^{10}$ GeV, with $a=b=1$ and $c=\kappa=0.1$:
\begin{itemize}

\item For leptogenesis scales $5\times10^8\,\,\mathrm{GeV}<T_{\rm lep}<T_{\rm RH}$, fermion mass mixing is $>T$ and fermionic modes are suppressed. Mixing from scalars gives:\\
 $T_{\rm lep}=10^8-10^{10}$ GeV and $m_X=5$ GeV.

\item Below $T_{\rm lep}=5\times10^8$ GeV:

\begin{enumerate}

\item $T_{\rm lep}=5\times10^8$ GeV and $m_X\sim5 $ GeV. Mixing is dominated by fermions because\\
$\mu_X\approx3\times10^8$ GeV $\gg m_{3/2}$.
\item $T_{\rm lep}=5\times10^7$ GeV and $m_X\sim5 $ GeV. Mixing is dominated by fermions because\\
$\mu_X \approx 3\times10^5$ GeV $> m_{3/2}$.

\end{enumerate}
Scalar mixing dominates for $T_{\rm lep}\lesssim10^7$ GeV and asymmetry transfer is suppressed, consistent with $X$ masses up to 100 TeV depending on the leptogenesis time. Note that, even though the gap between the reheat temperature and leptogenesis scale is wider in this example than for the low reheat case, the $\phi$ VEV (and hence mass mixing) is still large enough to keep the system in chemical  equilibrium at $T_{\rm lep} =5\times10^7$ GeV. The reason is that, with a higher temperature, the reheat time is also earlier and so the \emph{time} of leptogenesis is earlier than it would be with a lower reheat temperature. Since $\phi\sim 1/t$, this leads to a larger mixing.

\end{itemize}

To summarize our findings for heavy moduli, asymmetry transfer from higher-dimensional moduli couplings to $XN$ is consistent with dark matter masses ranging from 5 GeV to the weak scale over a range of reheat temperatures and leptogenesis scales. We generally expect scalar mixing to dominate when $m_{XN}>T$ and the fermion modes are suppressed. With lower $T_{\rm lep}$, fermion mixing dominates over scalar mixing when $m_{XN}$ is just below $T$, while scalar mixing dominates once again at low $T_{\rm lep}$ because $m_{XN}$ is small and the scalar mixing has an $m_{3/2}$ enhancement.
\\

Moving on to {\bf light moduli}, $\phi=\MPl$ at the time of reheating. Therefore, if leptogenesis occurs at or near the reheat temperature, the fermion $X$ and $N$ are thermally suppressed and the dominant asymmetry transfer happens between scalars. The scalar mixing is small because $m_{3/2}\sim m_\phi$ is small in this case, and the resulting transfer is out of equilibrium. In fact, for the points shown below, mixing from scalars alone cannot transfer enough asymmetry to be consistent with a dark matter mass below 100 TeV, which is the upper limit for fields whose symmetric components can be annihilated by perturbative couplings.

We present benchmark points for reheat temperatures $T_{\rm RH}=10^9$ GeV and $T_{\rm RH}=10^{11}$ GeV.  We begin with $T_{\rm RH}=10^9$ GeV and  $m_\phi = 3$ keV, which is the largest moduli mass consistent with decay and overclosure constraints.
\begin{itemize}

\item For leptogenesis scales $2\times10^6\,\,\mathrm{GeV}<T_{\rm lep}<T_{\rm RH}$, fermion mass mixing is $>T$ and fermionic modes are suppressed. Mixing from scalars is highly out-of-equilibrium and transfers no appreciable asymmetry.

\item For leptogenesis scales $T_{\rm lep}<2\times10^6\,\,\mathrm{GeV}$, fermion mixing is dominant. The asymmetry transferred to $X$ drops with $T_{\rm lep}$. Some benchmark points are
\begin{enumerate}
\item $T_{\rm lep}=2\times10^6$ GeV and $m_X=5$ GeV.
\item $T_{\rm lep}=5\times10^5$ GeV and $m_X=5$ GeV.
\item $T_{\rm lep}=10^5$ GeV and $m_X=160$ GeV.

\end{enumerate}
\end{itemize}
For the first two benchmark points, mixing is large and chemical equilibrium is attained as outlined in the previous paragraph. The transfer rate is suppressed for lower $T_{\rm lep}$ and so $\Delta_X \ll \Delta_{\rm B}$ for the last case. Higher $X$ masses are consistent with lower $T_{\rm lep}$, but in practice, it is very difficult to achieve leptogenesis at such a low scale.

For a high reheat temperature of $T_{\rm RH} = 10^{11}$ GeV, with $a=b=c=\kappa=1$. As before, we use $m_\phi = 3$ keV.
\begin{itemize}

\item For leptogenesis scales $8\times10^8\,\,\mathrm{GeV}<T_{\rm lep}<T_{\rm RH}$, fermion mass mixing is $>T$ and fermionic modes are suppressed. Mixing from scalars is highly out-of-equilibrium and transfers no appreciable asymmetry.

\item For leptogenesis scales $T_{\rm lep}<8\times10^8\,\,\mathrm{GeV}$, fermion mixing is dominant. The asymmetry drops with $T_{\rm lep}$. Some benchmark points are
\begin{enumerate}
\item $T_{\rm lep}=7\times10^8$ GeV and $m_X\sim5$ GeV.
\item $T_{\rm lep}=5\times10^8$ GeV and $m_X\sim5$ GeV.
\item $T_{\rm lep}=10^8$ GeV and $m_X=2.2$ TeV.
\item $T_{\rm lep}=3\times10^7$ GeV and $m_X=80$ TeV.

\end{enumerate}
\end{itemize}
For the first two points, $\Gamma_{N\rightarrow X}>H$ and the system is in chemical equilibrium, leading to $n_X\approx n_{\rm B}$ and $m_X\sim5$ GeV. For the latter two, the transfer rate is suppressed and $n_X\ll n_{\rm B}$, consistent with $X$ above the weak scale.
\\

There are two principal differences between the results for light and heavy moduli. Since the modulus mass is tied to $m_{3/2}$, the ratio of soft scales between light and heavy moduli is $\sim \mathrm{keV}/\mathrm{TeV}\sim10^{-9}$. As a result, scalar mixing is much more highly suppressed with light moduli. The second difference is that, for light moduli, $\phi$ only begins oscillation at reheating, whereas for heavy moduli oscillation begins before reheating. This means that the $\phi$ VEV is larger for light moduli, leading to larger fermion mixing.

For both light and heavy moduli in the limit $\Gamma_0\ll H$, the particles do not undergo any scatterings with the thermal background in a Hubble volume. This means that $X$ and $N$ remain in a coherent state given by the oscillation between the two states. The asymmetry transferred in this case is dependent only on the mixing angle. The requirement that $\Gamma_0\ll H$ is, however, at odds with the condition that the symmetric component of the dark matter be efficiently depleted. Therefore, the scenario of oscillation with no thermal scatterings is not a viable asymmetric dark matter model.

In summary, we have presented a number of scenarios where mixing between dark matter and lepton-number-carrying fields is induced by moduli VEVs. Depending on the reheat temperature and scale of leptogenesis, the mixing can be very rapid, bringing the two sectors into chemical equilibrium and transferring an $X$ asymmetry consistent with GeV-scale dark matter. Alternatively, the mixing can be suppressed, leading to dark matter at the weak scale or heavier. The mixing suppression can be very large, but the largest possible dark matter masses are $\sim100$ TeV, above which point the symmetric component of $X$ cannot be annihilated by perturbative couplings.

\subsubsection{Moduli cosmology}\label{sec:dilution}
Moduli are gravitationally coupled and  therefore long-lived, with characteristic decay widths
\ben\label{eq:moduliwidth}
\Gamma_\phi \approx \frac{m_\phi^3}{8\pi\MPl^2}.
\een
As a result, moduli tend to dominate the energy density of the universe until the time of their decay. The entropy accompanying moduli decay can greatly dilute any existing relics, including the baryon asymmetry. Furthermore, if moduli decay occurs after Big-Bang Nucleosynthesis (BBN), the resulting entropy generation alters the predictions for the light element abundances and such a scenario is in conflict with observations. There are two possibilities: either moduli decay reheats the universe to above the scale of BBN (about 5 MeV), or they are stable to the present day. These possibilities require heavy moduli and light moduli, respectively.

Assuming that the energy of the universe is moduli-dominated for the case of heavy moduli, the reheat temperature $T_1$ from moduli decay is given by
\ben
T_1 = \left(\frac{30m_\phi^2\,\phi(t=1/\Gamma_\phi)^2}{\pi^2g_*}\right)^{1/4}.
\een
Since $T_1 > T_{\rm BBN}\approx 5$ MeV, we determine that $m_\phi \gtrsim50$ TeV.

The temperature immediately before $\phi$ decay is
\ben
T_0 = T_{\rm RH}(t_{\rm RH}\Gamma_\phi)^{2/3}.
\een
The generated entropy (and hence the dilution of particle relics) is
\ben
\frac{s_1}{s_0} = \frac{T_1^3}{T_0^3}.
\een
For $m_\phi=100$ TeV and $T_{\rm RH} \sim 10^9$ GeV, the dilution factor is $\mathcal O(10^{10})$. This requires a primordial baryon asymmetry of $n_{\rm B}/s\sim1$ to give the correct abundances at late times.

One method of efficient baryogenesis is Affleck-Dine leptogenesis \cite{Affleck:1984fy,Dine:1995kz} along a flat direction with non-zero $N$ and Standard Model lepton number. If the entropy generated from the decay of the flat direction is larger than that from the inflaton, then $n_B/s$ can be of order unity \cite{Dine:1995kz}. The baryon asymmetry is
\ben
\frac{n_{\mathrm B}}{s} \approx \frac{n_{\mathrm B}}{n_{\rm FD}}\,\frac{T_{\rm RH}}{m_{\rm FD}}\left(\frac{m_{\rm FD}}{\MPl}\right)^{2/(n-2)},
\een
where $n_{\mathrm B}/n_{\rm FD}\sim\mathcal O(1)$ in the standard Affleck-Dine set-up. For $m_{\rm FD}\sim$ TeV, $T_{\rm RH}\sim10^9$ GeV, and $n\ge7$, then $n_{\mathrm B}/s\sim\mathcal O(1)$ and the dilution from moduli decay leads to the correct baryon asymmetry at late times. Examples of a flat direction  lifted at dimension-8 and invariant under the Standard Model and $\mathrm U(1)'$ gauge groups include: $\phi_{\rm FD}^4 = NLL\bar E$ and $N\bar U\bar D\bar D$, where we impose $R$-parity to forbid a lifting term of dimension-4. A detailed account of Affleck-Dine leptogenesis (baryogenesis) and moduli decay can be found in \cite{Kawasaki:2007yy}. Other, highly efficient, non-standard leptogenesis mechanisms (such as resonant leptogenesis \cite{Pilaftsis:2003gt}) are also possible, but less generic.

If the moduli are instead light,  they can be stable on cosmological scales. From cosmic ray data, the constraint on the dark matter lifetime is $\tau\gtrsim10^{26}$ seconds \cite{Meade:2009iu}. To be conservative, we require that the moduli also have a longer lifetime than this. Using the moduli decay rate (\ref{eq:moduliwidth}), long-lived moduli have masses $m_\phi \lesssim7$ MeV.

There are additional constraints on moduli masses from the requirement that moduli do not overclose the universe. If the initial $\phi$ VEV is $\MPl$ and oscillation begins at a temperature $T_{\rm osc}$, then $\phi$ energy density to entropy ratio is
\ben
\frac{\rho_\phi}{s} \approx \frac{m_\phi^2\MPl^2}{g_*T_{\rm osc}^3}.
\een
To avoid overclosing the universe, and so that moduli do not comprise a substantial fraction of the dark matter, we require that $\rho_\phi/s$ be less than the value for baryons, $\rho_{\rm B}/s \approx 10^{-10}$ GeV. If oscillation begins when $H(T_{\rm osc}) = m_\phi$, then $\phi$ overcloses the universe for $m_\phi >10^{-26}$ eV \cite{Conlon:2007gk}. Since we expect $m_{\phi}\sim m_{3/2}$, there are no viable models with such low scales of SUSY breaking.

If oscillation begins at earlier times, however,  the $\rho_\phi$ per comoving volume is suppressed and larger $\phi$ masses are allowed. One example of how this could happen is if the negative terms proportional to $H^2$ in the moduli potential (\ref{eq:modpotential})  arise only due to couplings with inflaton. If this is the case, the mass-squared of $\phi$ is driven positive when the inflaton decays and oscillation begins at the time of reheating, giving
\ben\label{eq:overclosure}
\frac{\rho_\phi}{s} \approx \frac{m_\phi^2\MPl^2}{g_*T_{\rm RH}^3}<\frac{\rho_{\rm B}}{s}.
\een
For example, if $m_\phi=0.1$ keV and $T_{\rm RH}=10^{10}$ GeV in the scenario where oscillation begins at reheating, the present-day moduli density is smaller than the baryon density.

\subsubsection{Mixing due to flat directions}
Combinations of fields with vanishing $\mathcal F-$ or $\mathcal D-$terms are also flat and can have large VEVs in the early universe. Such directions are typically lifted by higher-dimensional Planck-suppressed operators. For example, the potential of a flat direction lifted by a superpotential term of dimension 4 is
\ben
V(\phi) = (m^2-a^2H^2)|\phi|^2+\frac{c}{\MPl^2}|\phi|^6.
\een
The interplay between the negative mass-squared coming from the background energy and the $\phi^6$ lifting term gives $\phi$ a VEV for $H> m/a$,
\ben
\langle\phi\rangle = \left[\frac{\MPl^2(aH^2-m^2)}{3c}\right]^{1/4}.
\een
For $H\gg m/a$, $\langle\phi\rangle\sim H$. This is much smaller than the moduli VEV, which is typically $\MPl$. As a result, $m_{XN}$ is also suppressed relative to the moduli-induced mass mixing and the asymmetry transfer is out of equilibrium since it was only marginally so for much of the parameter space with mass mixing. The outcome is similar to the case for moduli when $T_{\rm lep}\ll T_{\mathrm c}$ and $m_{XN}$ was suppressed. This favors TeV scale or higher dark matter.

There are additional complications to flat direction VEVs giving mass mixing. The direction coupling to $XN$ cannot be comprised solely of MSSM fields, since the lowest allowed lifting operator would be dimension-4 in the superpotential and lead to rapid dark matter decay. The flat directions can also induce mass mixing between $X$ and other fields in the theory through renormalizable superpotential couplings, giving $X$ a very large mass and lifting it out of the theory. Such models typically require more complicated field content and symmetries to give large asymmetry transfer from $N\rightarrow X$. Although possible, the smaller field values make models more cumbersome so we focus on the simpler cases in this paper.

\subsection{Mixing induced by background energy}\label{sec:bkd}
An alternate, even more generic origin of mass mixing at early times is through couplings to fields dominating the energy density of the early universe.  We consider the specific case of couplings to kinetic terms of relativistic fermions $\psi$ in a thermal background.  The higher-dimensional operators are
\ben\label{eq:kinetic}
\Delta \mathcal L\supset \frac{c}{\MPl^2}\left(i\psi^{\dagger}\gamma^{\mu}D_{\mu}\psi\right)\left(XN+\mathrm{h.c.}\right).
\een
In a supersymmetric theory, such terms arise from higher-dimensional corrections to the K\"ahler potential.
If $\psi$ are fields dominating the background density, we have
\ben
\langle\psi_{\Sigma}^{\dagger}\gamma^{\mu}D_{\mu}\psi_{\Sigma}\rangle = \frac{\pi^2}{30}g_*T^4,
\een
where $g_*$ is the number of relativistic degrees of freedom. We have a mass mixing between scalars
\ben\label{eq:scalarmix2}
m_{XN}^2 = \frac{\pi^2\,c\,g_*\,T^4}{30\MPl^2} \approx \frac{\pi^2\,c}{82.8}\,H^2.
\een
With a higher-dimensional coupling (\ref{eq:kinetic}), the decay rate is suppressed by $(m_X/\MPl)^4$ and for $m_X\lesssim10^6$ GeV, the dark matter is stable.

Unlike the case of moduli-induced mixing, direct mixing of fermions by a kinetic term occurs only with operators of higher dimension than those inducing scalar mixing. We show below that the asymmetry transfer mechanism between scalars is viable only for very high temperatures (around the GUT scale), and this would be pushed even higher if the mixing were suppressed by an additional power of $T/\MPl$. Therefore,  it is only the scalar mixing through the thermal background  that is relevant.

We first estimate the expected parameters that give rise to a viable model and subsequently solve the density matrix evolution equations numerically. The total thermal interaction rate, averaged between $X$ and $N$, and using the couplings from Section \ref{sec:moduli} is approximately
\ben
\Gamma_0 \sim \frac{1}{8\pi^3}\left(g^4+\frac{1}{2}y^2\right)T,
\een
where $y$ and $g$ are Yukawa and gauge couplings given in (\ref{eq:couplings}). Because the mass mixing is proportional to $H$, it falls as $T^2$, and therefore asymmetry transfer is highest when $T$ is large and leptogenesis occurs at a high scale. Consequently, the out-of-equilibrium conditions and neutrino masses are consistent with $y\sim1$. We therefore expect the hierarchy $yT\gg m_{XN}\gg \mu$ to hold. The mass splittings and mixing angle in this limit are
\bean
\epsilon &\approx& \frac{y^2}{6}T,\\
\sin\theta &\approx& \frac{m_{XN}^2}{y^2\,T^2}.
\eean
For $g\sim1$, it is always true that $\Gamma_0\gg\epsilon$, and so (\ref{eq:asymtransfer}) reduces to
\ben
\Gamma_{N\rightarrow X^{\dagger}} = \frac{\epsilon^2}{\Gamma_0}\sin^2\theta\cos^2\theta\approx
\frac{2\pi^2\,T}{9(g^4+y^2/2)}\left(\frac{m_{XN}}{T}\right)^4.
\een
Integrating over a Hubble time and substituting (\ref{eq:scalarmix2}) gives
\ben
\frac{\Gamma_{N\rightarrow X^{\dagger}}}{H} \approx 0.07\pi^7\,g_*^{3/2}\left(\frac{T}{\MPl}\right)^3.
\een
For temperatures around $T_{\mathrm{GUT}}\sim10^{16}$ GeV, we obtain a fractional $X$ asymmetry (relative to lepton number) of around $5\times10^{-4}$, which would be the correct asymmetry for a few TeV dark matter candidate. In reality, the accumulated asymmetry is larger by a factor of $\sim5$ when we account for integrating over multiple Hubble times, as we find when we solve the full density matrix evolution equations.

We solve the system of equations (\ref{eq:rhoscalar}) numerically. We present three benchmark points giving $\Omega_{\mathrm{DM}}\approx5\Omega_{\mathrm b}$ with $y=1$ and $c=10$:

\begin{itemize}

\item $T_{\mathrm{lep}}=1.8\times10^{16}$ GeV, $g=1$ and $m_X=1$ TeV,

\item $T_{\mathrm{lep}}=1.8\times10^{16}$ GeV, $g=3$ and $m_X=10$ TeV,

\item $T_{\mathrm{lep}}=5\times10^{16}$ GeV, $g=10$ and $m_X = 50$ TeV.

\end{itemize}
 We have assumed that higher dimensional operators mix $X$ with all three generations of $N$. Asymmetries consistent with higher mass dark matter particles are also possible when $T_{\mathrm{lep}}<10^{16}$ GeV or for $g<1$, but in these cases, the couplings are insufficiently large to eliminate the symmetric component. Asymmetries consistent with lower masses of $X$ are also possible for higher reheat temperatures, but since $T_{\rm RH}$ is already borderline given the gravitino problem and the risk of reintroducing the flatness and monopole problems \cite{Bassett:2005xm}, we do not consider any higher reheat temperatures.

The temperature scales required for scalar mass mixing are higher than we may typically expect from a supersymmetric theory due to the gravitino problem. The gravitino problem can be alleviated in theories with very heavy gravitinos, where the gravitinos decay into axions \cite{Asaka:2000ew} or some other non-Standard Model particle. Alternatively, one may envision a non-supersymmetric theory with mixed scalars whose scales are stabilized by some unknown mechanism.

Note also that the asymmetry transferred by mass mixing is dominant over transfer from thermal scatterings from (\ref{eq:kinetic}). The rate of such processes is
\ben
\Gamma_{X\Sigma\rightarrow N\Sigma}\sim\frac{T^5}{\MPl^4} < H.
\een
Integrating the rate over a Hubble time gives $T^3/\MPl^3\sim10^{-7}$ even for the highest temperatures we consider, $T\sim10^{16}$ GeV and is therefore subdominant to the effect of mass mixing.

\subsection{Phenomenology}
Unlike the two Higgs model in Section \ref{sec:twost}, the minimum interactions necessary between dark matter and the visible sector for the models in this section are Planck-suppressed and there is no built-in mechanism for the annihilation of the symmetric component. As a result, the phenomenology is model-dependent.
The main hope of seeing such a model is through the mechanism for annihilation. In Sections \ref{sec:moduli} and \ref{sec:bkd},  we considered a $\mathrm U(1)'$ gauge interaction under which $X$ and $N$ are charged, although other possibilities exist. Direct detection bounds strongly constrain the strength of this interaction with quarks. As a result, signals would be small at hadron colliders.

As in Section \ref{sec:2higgspheno}, we do not expect asymmetric dark matter models to have large indirect detection signals. The mechanisms discussed there, however, could yield a non-zero anti-dark-matter population at late times, and so the model would not necessarily be ruled out by a confirmed indirect signal.

Since our model consists of $X$ mixing with a lepton-number-carrying gauge singlet, as is present in Dirac leptogenesis, the determination of the Dirac or Majorana nature of the neutrino masses can give insight into the components of our model. However, if neutrinos are discovered to have Majorana masses, this will not rule out the models listed above but would favor a more complex model than the minimal Dirac scenario or suggest Affleck-Dine mechanism for generating original asymmetries.

\section{Conclusions}
The global symmetries of the universe could have been very different in the early universe than they are today. In particular, independent dark matter and baryon asymmetries at late times can emerge from a common symmetry at early times and explain the baryon-dark matter near-coincidence.  In this paper, we show how large field backgrounds in the early universe can induce mass mixings between dark matter and visible matter, sometimes leading to distinctive relations between the dark matter and baryon densities that accommodate a wide range of dark matter masses. Such mixing can arise in renormalizable Lagrangian terms with field VEVs linked to scales already present in the theory, such as the scale of electroweak symmetry breaking. In this case, the dark matter mass is naturally either $\sim$ GeV or in the range $300-500$ GeV. More generically, mixing can also arise from Planck-suppressed operators, which we expect to break the global symmetries present at low scales. This leads to mixing even in scenarios where there is no renormalizable connection between dark and visible matter fields. In this case, the transferred asymmetry is consistent with dark matter masses above 1 GeV, with weak scale masses preferred in the case of moduli-induced mixing and leptogenesis at the intermediate scale or in the case of mixing due to background energy when leptogenesis happens at the GUT scale. The dark matter masses are typically higher for
lower leptogenesis scales, with the only upper bound on the dark matter masses being the requirement that the symmetric component be annihilated by perturbative couplings.

We summarize our results for the various cases we consider in the table below.
\begin{itemize}
\item Two-stage phase transition:\\
\begin{tabular}{|c|c|c|c|}
  \hline
            & Relativistic $X$  & Thermal-dominant $X$ & Mixing-angle-dominant $X$ \\\hline
  Fermion DM mass & $\mathcal O(\rm GeV)$ & 300-500\,\,\rm GeV & 400-500\,\,\rm GeV \\
  \hline
\end{tabular}\label{2higgstable}
\item Moduli-induced mixing:\\
\begin{tabular}{|p{2.6cm}|p{5.6cm}|p{6.6cm}|}
  \hline
                 & \begin{center}{\bf Heavy moduli}\end{center}  & \begin{center}{\bf Light stable moduli}\end{center}\\\hline
  DM mass with scalar mixing & $T_{\rm RH}\sim10^8\,\,\rm GeV$: $90$ GeV$-100$ TeV; $T_{\rm RH}\sim10^{10}\,\rm GeV$: $\sim \mathcal O(\rm GeV)-100\rm\,\,TeV$ & Transfer highly suppressed\\\hline
  DM mass with fermion mixing& $T_{\rm RH}\sim10^8-10^{10}\,\,\rm GeV$: $\mathcal O(\rm GeV)$ & $T_{\rm RH}\sim10^9\,\rm GeV$: $\sim\mathcal O(\rm GeV)$$-100$ TeV; $T_{\rm RH}\sim10^{11}\,\rm GeV$: $\sim\mathcal O(\rm GeV)$$-100 \rm \,\,TeV$
   \\
  \hline
\end{tabular}
\item Background energy induced mixing:\\
\begin{tabular}{|c|c|}
  \hline
  DM mass with scalar mixing & $T_{\rm RH}\sim10^{16}\,\,\rm GeV$ : $1-100\,\,\rm TeV$ \\
  \hline
\end{tabular}

\end{itemize}

This work expands the range of possible transfer mechanisms in asymmetric dark matter models and provides new considerations for dark matter model building. We see that wide range of dark matter models and broad range of masses is possible. It is important to explore
such possibilities and possible means to distinguish among these and other models
so we can eventually learn the true nature of dark matter.

\section*{Acknowledgements}
We wish to thank Jeffrey Lyons for the NYFC challenge. We also thank Howard Georgi, Dilani Kahawala, Tongyan Lin, Ann Nelson, and David Simmons-Duffin for helpful conversations. YC is supported by NSF grant number PHY-0855591 and the
Harvard Center for the Fundamental Laws of Nature. LR is supported by NSF grant number PHY-0855591.

\appendix
\section*{Appendices}
\section{Two Higgs model vacua}\label{app:twohiggs}
\subsection{Vacua and stability}

The vacua of the potential can be determined by finding the extrema of the potential. Here, we follow the analysis of \cite{Land:1992sm}. We perform the analysis at finite temperature and then discuss the particular case of the zero temperature vacua.

Extremizing the potential gives
\bean
\langle h\rangle &=& 0,\,\,\,\frac{1}{2}\sqrt{\frac{4\mu_1^2-2k_3\langle\phi\rangle^2-\alpha_1T^2}{k_1}},\\
\langle\phi\rangle &=&0,\,\,\,\frac{1}{2}\sqrt{\frac{4\mu_2^2-2k_3\langle h\rangle^2-\alpha_2T^2}{k_2}}.
\eean
At a given temperature, there are four distinct critical points of the potential:
\begin{enumerate}
\item\ben
\langle h\rangle = \langle\phi\rangle = 0,
\een
\item\ben\label{eq:phivev}
\langle h\rangle = 0,\,\,\, \langle\phi\rangle = \frac{1}{2}\sqrt{\frac{4\mu_2^2-\alpha_2T^2}{k_2}},
\een
\item\ben
\langle \phi\rangle = 0,\,\,\, \langle h\rangle = \frac{1}{2}\sqrt{\frac{4\mu_1^2-\alpha_1T^2}{k_1}},
\een
\item\ben \label{eq:sadpt}
\langle h\rangle = \sqrt{\frac{8k_2\mu_1^2-4k_3\mu_2^2+(\alpha_2k_3-2\alpha_1k_2)T^2}{8k_1k_2-2k_3^2}},\,\,\, \langle\phi\rangle =\sqrt{\frac{8k_1\mu_2^2-4k_3\mu_1^2+(\alpha_1k_3-2\alpha_2k_1)T^2}{8k_1k_2-2k_3^2}}.
\een
\end{enumerate}
For extrema 2 and 3, the vacuum is only well-defined if
\ben
T^2 < T_{\mathrm c i}^2 \equiv \frac{4\mu_i^2}{\alpha_i}.
\een
These temperatures are sufficiently important that we label them \emph{critical temperatures} and they indicate the point at which the fields $h$ and $\phi$ can condense.

We now examine the stability of the extrema of the potential.
\begin{enumerate}

\item The symmetric vacuum is only stable for $T^2 > \mathrm{max}(T_{\mathrm c1}^2,T_{\mathrm c2}^2)$. Below these temperatures, the negative mass-squared parameter in the Lagrangian dominates and tends to drive the fields to non-zero values. At high temperature, the scalars begin in the symmetric phase and later condense when $T$ passes below one of the threshold temperatures.

\item The vacuum where $\langle h\rangle=0$ and $\langle\phi\rangle\neq0$ is stable given $T<T_{\mathrm c2}$ and $k_3\alpha_2(T_{\mathrm c2}^2-T^2) > 2k_2\alpha_1(T_{\mathrm c1}^2-T^2)$. For $T_{\mathrm c2}>T>T_{\mathrm c1}$, this condition is automatically satisfied. If $T<T_{\mathrm c1}$ and $T_{\mathrm c2}$, then the vacuum is only stable down to the temperature
\ben \label{eq:to}
T_0^2 = \frac{4(2k_2\mu_1^2-k_3\mu_2^2)}{k_2\alpha_1-k_3\alpha_2},
\een
at which point this extremum becomes a saddle point.

\item The vacuum where $\langle \phi\rangle=0$ and $\langle h\rangle\neq0$ is exactly the same as the one in vacuum \#2, except with indices $1\leftrightarrow2$.

\item Finally, the conditions for a stable vacuum at both $h$ and $\phi$ non-zero are
\bean \label{eq:sadpt1}
2k_2\alpha_1(T_{\mathrm c1}^2-T^2)-k_3\alpha_2(T_{\mathrm c2}^2-T^2) &>&0,\\
2k_1\alpha_2(T_{\mathrm c2}^2-T^2)-k_3\alpha_1(T_{\mathrm c1}^2-T^2) &>&0,\\\label{eq:sadpt2}
4k_1k_2-k_3^2 &>&0.\label{eq:sadpt3}
\eean
It is not immediately apparent whether this is stable or not, but we demonstrate below out that for all parameters of interest to us (namely those that give a two stage phase transition), the conditions are be violated and this is be a saddle point.
\end{enumerate}

Because extremum \# 4 is a saddle point, there is only one non-zero VEV in any vacuum. This gives us two options: the first is that the field that initially condenses is also the field with a VEV at zero temperature (we want this to be $h$) and hence there is only a one stage phase transition. The other possibility is that the initial field to condense ($\phi$) has a higher zero-temperature energy when condensed than the vacuum with non-zero $h$, and so at late times there is a phase transition where the $\phi$ VEV shuts off and the $h$ VEV is turned on. Because there is a classical energy barrier between these two minima of the potential, this generally leads to a first-order phase transition (see Appendix \ref{app:bubble}).

\subsection{Parameter constraints}\label{app:paramconstraints}
In this section, we determine the parameters that lead to a viable two stage phase transition. We begin by matching the parameters onto the known zero temperature physics. In particular, we would like $\langle\phi\rangle=0$ and $\langle h\rangle = v$ at zero temperature, where $v=$ 246 GeV. To ensure that $h$ condenses instead of $\phi$, we require the vacuum energy in the state where $h$ is condensed to have a lower energy than the state where $\phi$ is condensed. The vacuum energy in the $\langle\phi\rangle=0$ vacuum at zero temperature is
\ben
V_1(T=0) = -\frac{\mu_1^4}{k_1}
\een
and similarly for the $\langle h\rangle=0$ vacuum energy $V_2$. For the former to be lower, we require $|V_1(T=0)|>|V_2(T=0)|$ or, alternately,
\ben \label{eq:cons1}
\frac{\mu_1^4}{\mu_2^4} > \frac{k_1}{k_2}.
\een
Imposing this constraint, and therefore being in the $\langle\phi\rangle=0$ vacuum at zero temperature, we can determine the physical Higgs mass from the quadratic term,
\ben
m_h^2 = 8\mu_1^2.
\een
The zero temperature Higgs mass has not yet been determined, but electroweak fits prefer a value as close as possible to the LEP bound of 115 GeV, so we choose $m_H = 120$ GeV. We can then determine the parameters $\mu_1$ and $k_1$ using
\bean
\mu_1 &=& \frac{m_h}{\sqrt 8}\approx 42.4\,\,\mathrm{GeV},\label{eq:mu1}\\
k_1 &=& \frac{1}{8}\left(\frac{m_h}{v}\right)^2\approx 0.0297.
\eean
Meanwhile, $\phi$ gets a mass due to the $\phi^2 h^2$ term, and its value is
\ben
m_{\phi}^2 = 2\mu_1^2\frac{k_3}{k_1}-4\mu_2^2.
\een
This gives us our second constraint,
\ben \label{eq:cons2}
k_3 > \frac{2\mu_2^2}{\mu_1^2}k_1,
\een
which requires that the mixed quartic term be large enough to compensate for the negative $\phi$ mass-squared term. If (\ref{eq:cons2}) is not satisfied, then $\phi$ is also driven to a non-zero value (vacuum \#4 above).

In order to have a two stage phase transition, we would like the system to first condense in the $\phi$ direction. This imposes the requirement that $T_{\mathrm c1} < T_{\mathrm c2}$, which can be expressed as
\ben \label{eq:cons3}
\frac{\mu_1^2}{\mu_2^2}<\frac{\alpha_1}{\alpha_2};
\een
combining this with (\ref{eq:cons1}) gives
\ben
\sqrt{\frac{k_1}{k_2}} < \frac{\mu_1^2}{\mu_2^2} < \frac{\alpha_1}{\alpha_2}.
\een
Thus, we see that the mass terms are constrained to be roughly close in magnitude to one another in order to have a two stage phase transition.

\section{The two Higgs phase transition and bubble nucleation}\label{app:bubble}
\subsection{Tunneling and bubble nucleation}
There exists a potential barrier between the true and false vacua in our two Higgs model. The vacua become degenerate at the temperature
\ben\label{eq:degenerate}
T_{\mathrm d}^2 = \frac{4\left(\sqrt{k_2}\mu_1^2-\sqrt{k_1}\mu_2^2\right)}{\sqrt{k_2}\alpha_1-\sqrt{k_1}\alpha_2}.
\een
Below this temperature, the system can transition from the false vacuum to the true vacuum. The phase transition between the two vacua, should it occur, satisfies the following conditions:
\begin{enumerate}

\item It occurs at the very latest before big bang nucleosynthesis ($T\gg 10$ MeV) to satisfy observational constraints.

\item The $\phi$ VEV must turn off sufficiently rapidly that the asymmetry relation from mass mixing is not disturbed. This means either that the fields directly tunnel into the $\phi=0$ vacuum, or in the case where the system tunnels to a non-zero $\phi$ vev, that $\phi$ is rapidly damped to zero (on a time shorter than the rate of asymmetry transfer) following tunneling.

\item The expansion of the bubble wall should be faster than the asymmetry transfer rate from thermal scattering.

\end{enumerate}

To begin, we calculate the temperature at which bubble nucleation occurs. At large temperature, when the bubble wall $r_{\mathrm w}\gg \beta$, we can take the bubble solution to be constant in the period Euclidean time, $\tau$. The bounce solution is then given by the $\mathrm O(3)$ symmetric function minimizing the Euclidean action
\ben\label{eq:3daction}
S(T,\phi,h) = \frac{4\pi}{T}\int dr\,r^2\left[\frac{1}{2}\left(\frac{d\phi}{dr}\right)^2+\frac{1}{2}\left(\frac{dh}{dr}\right)^2+V(\phi,h,T)\right].
\een
The semi-classical probability $P$ of nucleating a bubble inside of a casual volume during the temperature interval $dT$ is \cite{Anderson:1991zb}
\ben
P \approx16\,\omega\,\xi^4\frac{\MPl^4}{T^5}\,e^{-S(T)}\,\Delta T,
\een
where $\omega$ is a dimensionless coefficient given by functional determinants from the path integral and $\xi\approx 1/34$ is a numerical factor coming from the Hubble constant. The result is dominated by the action $S$ in the exponential, making the precise value of $\omega$ irrelevant.  Taking $\omega\sim1$, $T\sim30$ GeV and $\Delta T\sim10$ GeV (for reasons that will be shown later), we find that $S\approx140$ gives a probability of nucleation equal to one. We take this as our benchmark for determining when nucleation occurs.

It is difficult to find the bounce solution to (\ref{eq:3daction}) numerically since there are two fields involved. Because we are simply interested in an estimate of the action, we instead make an ansatz that eliminates one of the fields. We take $h = f(\phi)$, with $f$ to be determined from $V(\phi,h)$, giving
\ben
S(T,\phi) = \frac{4\pi}{T}\int dr\,r^2\left[\frac{1}{2}\left(1+f'(\phi)^2\right)\left(\frac{d\phi}{dr}\right)^2+V(\phi,T)\right].
\een
The resulting equation of motion is
\ben
\frac{d^2\phi}{dr^2}+\frac{2}{r}\,\frac{d\phi}{dr} = \frac{1}{1+f'(\phi)^2}\left[\frac{\ptl V}{\ptl\phi} - \left(\frac{d\phi}{dr}\right)^2f'(\phi)\,f''(\phi)\right].
\een
The initial condition $\phi(0)$ is chosen to satisfy the bounce boundary conditions $\phi'(0)=0$ and $\phi(\infty)= \phi_{\mathrm{false}}$. This equation can be solved numerically using the ``overshoot-undershoot'' method \cite{Coleman:1977py}.

The ansatz $h=f(\phi)$ is chosen as the curve in $\phi-h$ space such that the smallest possible potential barrier is crossed while tunneling. We determine this curve by slicing the potential along lines of constant $\phi$ and determining the value of $h$ that minimizes $V$. While we cannot guarantee that this is the minimum action, it is plausible that it is among the curves that gives the smallest action, and it gives an upper bound on the action. If we find rapid bubble nucleation from our ansatz, then bubble nucleation \emph{does} occur, even if the exact instanton configuration is slightly different from the one determined here.

Bubble nucleation does not occur for all choices of parameters; for a particular choice of $k_3$, $k_2$ must be chosen sufficiently large to reduce the barrier between vacua. Numerical solutions are presented in Section \ref{sec:numerics}.

To address point \#2 above, we find that tunneling does not, in fact, take $\phi$ to zero but instead to some non-zero value. $\phi$ then undergoes damped oscillations about its minimum, $\phi=0$. The damping due to the Hubble constant, $H\sim10^{-13}$ GeV, is negligible. The dominant source of damping is decays of $\phi\rightarrow X+L$ and the conjugate process. These decays do not change the asymmetry in the $X$ or $L$ sectors, since it is $CP$-conserving. We require that $\Gamma_{\phi} \gg \Gamma_{L\rightarrow X}$ to avoid eliminating the asymmetry in $X$ by the residual mass mixing due to non-zero $\phi$. We can estimate $\Gamma_{\phi}$ by
\ben
\Gamma_{\phi} \sim \frac{y_X^2}{4\pi}\,m_{\phi}\,\sqrt{1-\frac{m_X^2}{m_{\phi}^2}}.
\een
The asymmetry washout term is \ref{eq:transferrate}, with $\Gamma_0$ given by the larger of
\bean
\Gamma_X &\sim& \frac{y_X^4}{8\pi^3}\,\frac{m_X^2}{m_{\phi}^4}\,T^3,\\
\Gamma_L &\sim& \frac{g_{\rm w}^4}{8\pi^3}T.
\eean
The first interaction comes from $XX\rightarrow LL$ through an intermediate $\phi$, while the second is $L^0$ interacting with $W$ bosons, whose mass is given dominantly by the $h$ VEV. For a benchmark point  with $y_X=1$, $m_X=400$ GeV, $m_{\phi}=450$ GeV and $T_{\rm N}=60$ GeV (as is a typical scale for bubble nucleation), we get $\Gamma_{\phi}=16$ GeV while $\Gamma_0\approx\Gamma_L=7.6\times10^{-2}$ GeV. Thus, the VEV of $\phi$ safely damps away before the asymmetry relation from mass mixing is altered.

Finally, the bubble wall must sweep over a particle without it undergoing any thermal interactions. There is approximately one particle in a region with length $1/T$. The bubble wall takes a time $t_{\mathrm w}=(2\delta r+1/T)/v_{\mathrm w}$ to pass over the cube, where $\delta r$ is the bubble wall thickness and $v_{\mathrm w}$ is the velocity. This time must satisfy $t_{\mathrm w}\ll 1/\Gamma_{L\rightarrow X}$. We find that the wall thickness is always $\lesssim1\,\,\mathrm{GeV}^{-1}$. Using the rates of interaction in the false vacuum phase and assuming a small mixing angle, we determine that $v_{\mathrm w}>10^{-3}$ for the bubble wall passage to be out of equilibrium. This is easily satisfied for variants of the Standard Model with a first-order phase transition (the drag from top quarks in that case is similar to the drag for this bubble wall) \cite{Megevand:2009gh}. Therefore, the transition is strongly first-order and our instantaneous approximation for the state projection is valid.

For GeV-scale dark matter, where the mixing angle is large, there is no thermal suppression of any of the modes ($n_{\bar X'}=n_{L'}$) and neither the number densities nor the asymmetries change during the phase transition.

\subsection{Back-reaction on phase transition}

So far, we have neglected the back-reaction of the mass terms on the evolution of the bubble expansion at the first-order phase transition. In the limit of $T_{\mathrm t} \ll M$, the dark matter energy density present after the phase transition is entirely given by the repopulation of the heavy states due to the mass mixing. We must also concern ourselves with the ``drag'' associated with the top quarks becoming massive at the phase transition.

The energy density of the light mass eigenstates that become heavy $X$ fields following the phase transition is given by
\ben
\rho_X = m_X\sin^2\theta n_{L'} = \frac{2}{\pi^2}\sin^2\theta m_X T_{\mathrm N}^3,
\een
while the mass energy density of the top quarks following the phase transition is
\ben
\rho_t = m_t n_t = \frac{12}{\pi^2}\left(\frac{v_h(T_{\mathrm N})}{246\,\,\mathrm{GeV}}\right)m_t T_{\mathrm N}^3.
\een
The contribution from  top quarks dominates over the contribution from $X$ even for the largest viable masses for our model $m_X\sim500$ GeV, since there are six top quark degrees of freedom, giving an ``effective'' top quark mass of $6m_t \sim $ TeV.

At high temperature, the quarks have a large kinetic energy and this can be used to offset the vacuum energy expended to impart mass to the quarks. The energy density from the kinetic energy of the quarks (using $\langle p\rangle \approx 3T$) is
\ben
\rho_{\mathrm k} = \frac{36}{\pi^2}T_{\mathrm N}^4,
\een
and so for $T_{\mathrm t} > 50$ GeV, we find that the mass of the top quarks can be accounted for entirely from the kinetic energy and so this does not affect the progress of the phase transition.

In the region 30 GeV $< T_{\mathrm t} < $ 50 GeV, the situation is slightly more subtle. The energy density that goes into creating masses for the top quarks is comparable to the energy difference between the vacua immediately after $T_{\mathrm t}$. We must therefore not use $T_{\mathrm t}$ as the tunneling temperature and use instead the temperature at which the energy difference between the vacua is equal to the energy density from top quark masses. For a benchmark point with $y_X = 0.4$, $k_2=0.5$, $k_3=0.7$ and $\mu_2 = 80$ GeV, we find that $T_{\mathrm N} = 34$ GeV, whereas the temperature where the energy difference between vacua is $\rho_t$ is $T=30$ GeV, corresponding to an approximately 10\% difference. Similar results are found for nearby benchmark points.

Bubble nucleation rarely occurs below 30 GeV, so we do not consider this range further.

Fortunately, this does not significantly alter our results, since in the mass mixing-dominated limit, the final $X$ density is fixed only by $\tan^2\theta$, which has only a weak dependence on $T$. In fact, this gives us slightly more flexibility as the temperature for bubble nucleation can be about 10\% lower than before, giving more thermal suppression and a broader range of parameter space where the $X$ density is dominated by mass mixing.

\section{Review of Dirac leptogenesis}\label{app:Diraclepto}
We review a mechanism for realistic Dirac leptogenesis \cite{Dick:1999je}. As the universe cools below the temperature $T=M_{\psi}$, $\psi$ and $\bar\psi$ decay in two ways:
\bean
\psi&\rightarrow& N^{\dagger} H_u^{\dagger},\\
\psi &\rightarrow& L\chi;\\
\bar\psi &\rightarrow& N\,H_u,\\
\bar\psi &\rightarrow & L^{\dagger}\chi^{\dagger}.
\eean
The total decay rates of $\psi$ and $\bar\psi$ are equal, but they can decay with different rates into $N$ and $L$. This can give rise to an asymmetry of $L$ over $L^{\dagger}$. Because total lepton number is conserved by these interactions (assuming $\psi$ carries $L=1$), this asymmetry is equal and opposite to the asymmetry of $N$ over $N^{\dagger}$. Thus, there is an $N$ asymmetry that is directly correlated with the Standard Model lepton asymmetry (and hence related to the baryon asymmetry by sphaleron processes). $N$ and $L$ ultimately come into equilibrium due to oscillations once $H_u$ and $\chi$ have VEVs (see \cite{Manohar:1986gj}).

Eventually, $N$ and $L$ come into thermal equilibrium and the lepton asymmetry is destroyed at the time of equilibriation. This must happen after the time of BBN, since strong constraints exist on the Standard Model lepton asymmetry in that era. Fortunately, this equilibriation rate is suppressed by neutrino masses and is not important until $T\sim$ eV. We can see that this is true below $M_{\psi}$ by integrating out $\psi$, giving an effective superpotential
\ben
W_{\mathrm{eff}} \supset \frac{y_L\,y_N}{M_{\psi}} NH_u L\chi.
\een
At low temperatures, $\chi$ and $H_u$ have weak scale VEVs (which come about due to interactions with the SUSY breaking sector), giving neutrino masses
\ben
m_{\nu}\sim\frac{y_L\,y_N\,v\langle\chi\rangle}{M_{\psi}} < \mathrm{eV},
\een
meaning that $y_L\,y_N\sim0.1-1$ for $M_{\psi}\sim 10^{15}$ GeV and weak scale $\langle\chi\rangle$, while it is $y_L\,y_N\sim10^{-5}$ for $M_{\psi}\sim10^{10}$ GeV.


\begin{thebibliography}{99}


\bibitem{Kaplan:2009ag}
  D.~E.~Kaplan, M.~A.~Luty, K.~M.~Zurek,
  Phys.\ Rev.\  {\bf D79}, 115016 (2009).
  [arXiv:0901.4117 [hep-ph]].

\bibitem{An:2009vq}
  H.~An, S.~-L.~Chen, R.~N.~Mohapatra, Y.~Zhang,
  JHEP {\bf 1003}, 124 (2010).
  [arXiv:0911.4463 [hep-ph]];
  E.~J.~Chun,
  Phys.\ Rev.\  {\bf D83}, 053004 (2011).
  [arXiv:1009.0983 [hep-ph]];
  P.~-H.~Gu, M.~Lindner, U.~Sarkar, X.~Zhang,
  [arXiv:1009.2690 [hep-ph]];
  A.~Falkowski, J.~T.~Ruderman, T.~Volansky,
  [arXiv:1101.4936 [hep-ph]];
  Z.~Kang, J.~Li, T.~Li, T.~Liu, J.~Yang,
  [arXiv:1102.5644 [hep-ph]];
  D.~E.~Kaplan, G.~Z.~Krnjaic, K.~R.~Rehermann, C.~M.~Wells,
  [arXiv:1105.2073 [hep-ph]].

   \bibitem{Allahverdi:2010rh}
  R.~Allahverdi, B.~Dutta, K.~Sinha,
  Phys.\ Rev.\  {\bf D83}, 083502 (2011).
  [arXiv:1011.1286 [hep-ph]];

  \bibitem{Bell:2011tn}
  N.~F.~Bell, K.~Petraki, I.~M.~Shoemaker, R.~R.~Volkas,
   [arXiv:1105.3730 [hep-ph]];
  C.~Cheung, K.~M.~Zurek,
    [arXiv:1105.4612 [hep-ph]].

\bibitem{Nussinov:1985xr}
  S.~Nussinov,
  Phys.\ Lett.\  {\bf B165}, 55 (1985);
  S.~Dodelson, L.~M.~Widrow,
  Phys.\ Rev.\ Lett.\  {\bf 64}, 340-343 (1990).
  S.~M.~Barr, R.~S.~Chivukula, E.~Farhi,
  Phys.\ Lett.\  {\bf B241}, 387-391 (1990);
  S.~M.~Barr,
  Phys.\ Rev.\  {\bf D44}, 3062-3066 (1991);
  D.~B.~Kaplan,
  Phys.\ Rev.\ Lett.\  {\bf 68}, 741-743 (1992);
  D.~Hooper, J.~March-Russell, S.~M.~West,
  Phys.\ Lett.\  {\bf B605}, 228-236 (2005).
  [hep-ph/0410114];
  S.~B.~Gudnason, C.~Kouvaris, F.~Sannino,
  Phys.\ Rev.\  {\bf D74}, 095008 (2006).
  [hep-ph/0608055].

\bibitem{Cohen:2009fz}
  T.~Cohen, K.~M.~Zurek,
  Phys.\ Rev.\ Lett.\  {\bf 104}, 101301 (2010).
  [arXiv:0909.2035 [hep-ph]].

\bibitem{Cohen:2010kn}
  T.~Cohen, D.~J.~Phalen, A.~Pierce, K.~M.~Zurek,
  Phys.\ Rev.\  {\bf D82}, 056001 (2010).
  [arXiv:1005.1655 [hep-ph]];
  P.~-H.~Gu,
  Phys.\ Rev.\  {\bf D81}, 095002 (2010).
  [arXiv:1001.1341 [hep-ph]];
  J.~Shelton, K.~M.~Zurek,
  Phys.\ Rev.\  {\bf D82}, 123512 (2010).
  [arXiv:1008.1997 [hep-ph]];
  H.~Davoudiasl, D.~E.~Morrissey, K.~Sigurdson, S.~Tulin,
  Phys.\ Rev.\ Lett.\  {\bf 105}, 211304 (2010).
  [arXiv:1008.2399 [hep-ph]];
  N.~Haba, S.~Matsumoto,
  [arXiv:1008.2487 [hep-ph]];
  M.~Blennow, B.~Dasgupta, E.~Fernandez-Martinez, N.~Rius,
  JHEP {\bf 1103}, 014 (2011).
  [arXiv:1009.3159 [hep-ph]];
  J.~McDonald,
  [arXiv:1009.3227 [hep-ph]];
  L.~J.~Hall, J.~March-Russell, S.~M.~West,
  [arXiv:1010.0245 [hep-ph]];
  B.~Dutta, J.~Kumar,
  Phys.\ Lett.\  {\bf B699}, 364-367 (2011).
  [arXiv:1012.1341 [hep-ph]];
  M.~T.~Frandsen, S.~Sarkar, K.~Schmidt-Hoberg,
  [arXiv:1103.4350 [hep-ph]];
  J.~March-Russell, M.~McCullough,
  [arXiv:1106.4319 [hep-ph]].


\bibitem{Buckley:2010ui}
  M.~R.~Buckley, L.~Randall,
  [arXiv:1009.0270 [hep-ph]].

\bibitem{Weinberg:1974hy}
  S.~Weinberg,
  Phys.\ Rev.\  {\bf D9}, 3357-3378 (1974).

\bibitem{Land:1992sm}
  D.~Land, E.~D.~Carlson,
  Phys.\ Lett.\  {\bf B292}, 107-112 (1992).
  [hep-ph/9208227];
  A.~Hammerschmitt, J.~Kripfganz, M.~G.~Schmidt,
  Z.\ Phys.\  {\bf C64}, 105-110 (1994).
  [hep-ph/9404272].

\bibitem{Linde:1991km}
  A.~D.~Linde,
  Phys.\ Lett.\  {\bf B259}, 38-47 (1991);
  A.~R.~Liddle, D.~H.~Lyth,
  Phys.\ Rept.\  {\bf 231}, 1-105 (1993).
  [astro-ph/9303019];
    A.~D.~Linde,
  Phys.\ Rev.\  {\bf D49}, 748-754 (1994).
  [astro-ph/9307002];
  E.~J.~Copeland, A.~R.~Liddle, D.~H.~Lyth, E.~D.~Stewart, D.~Wands,
  Phys.\ Rev.\  {\bf D49}, 6410-6433 (1994).
  [astro-ph/9401011];
  E.~D.~Stewart,
  Phys.\ Lett.\  {\bf B345}, 414-415 (1995).
  [astro-ph/9407040];
  L.~Randall, M.~Soljacic, A.~H.~Guth,
  Nucl.\ Phys.\  {\bf B472}, 377-408 (1996).
  [hep-ph/9512439].

\bibitem{Manohar:1986gj}
  A.~Manohar,
  Phys.\ Lett.\  {\bf 186B}, 370 (1987).

\bibitem{Harvey:1990qw}
  J.~A.~Harvey, M.~S.~Turner,
  Phys.\ Rev.\  {\bf D42}, 3344-3349 (1990).

\bibitem{Abbiendi:2003ji}
  G.~Abbiendi {\it et al.} [ OPAL Collaboration ],
  Eur.\ Phys.\ J.\  {\bf C32}, 453-473 (2004).
  [hep-ex/0309014].

\bibitem{Raby:1987nb}
  S.~A.~Raby, G.~West,
  Nucl.\ Phys.\  {\bf B292}, 793 (1987).

\bibitem{Kopp:2009et}
  J.~Kopp, V.~Niro, T.~Schwetz, J.~Zupan,
  Phys.\ Rev.\  {\bf D80}, 083502 (2009).
  [arXiv:0907.3159 [hep-ph]].

\bibitem{Bityukov:1997ck}
  S.~I.~Bityukov, N.~V.~Krasnikov,
  Phys.\ Atom.\ Nucl.\  {\bf 62}, 1213-1225 (1999).
  [hep-ph/9712358].


\bibitem{Graesser:2011wi}
  M.~L.~Graesser, I.~M.~Shoemaker, L.~Vecchi,
   [arXiv:1103.2771 [hep-ph]].

\bibitem{Dick:1999je}
  K.~Dick, M.~Lindner, M.~Ratz, D.~Wright,
  Phys.\ Rev.\ Lett.\  {\bf 84}, 4039-4042 (2000).
  [hep-ph/9907562];
  H.~Murayama, A.~Pierce,
  Phys.\ Rev.\ Lett.\  {\bf 89}, 271601 (2002).
  [hep-ph/0206177].

\bibitem{Gelmini:1994az}
  G.~Gelmini, E.~Roulet,
  Rept.\ Prog.\ Phys.\  {\bf 58}, 1207-1266 (1995).
  [hep-ph/9412278].

\bibitem{Foot:1995qk}
  R.~Foot, M.~J.~Thomson, R.~R.~Volkas,
  Phys.\ Rev.\  {\bf D53}, 5349-5353 (1996).
  [hep-ph/9509327].

\bibitem{Dolgov:2003sg}
  A.~D.~Dolgov, F.~L.~Villante,
  Nucl.\ Phys.\  {\bf B679}, 261-298 (2004).
  [hep-ph/0308083].

\bibitem{Meade:2009iu}
  P.~Meade, M.~Papucci, A.~Strumia and T.~Volansky,
  Nucl.\ Phys.\  B {\bf 831}, 178 (2010)
  [arXiv:0905.0480 [hep-ph]].

\bibitem{Dine:1995kz}
  M.~Dine, L.~Randall, S.~D.~Thomas,
  Nucl.\ Phys.\  {\bf B458}, 291-326 (1996).
  [hep-ph/9507453].


\bibitem{Kolb:1990vq}
  E.~W.~Kolb, M.~S.~Turner,
  Front.\ Phys.\  {\bf 69}, 1-547 (1990).

\bibitem{Affleck:1984fy}
  I.~Affleck, M.~Dine,
  Nucl.\ Phys.\  {\bf B249}, 361 (1985).

\bibitem{Kawasaki:2007yy}
  M.~Kawasaki, K.~Nakayama,
  Phys.\ Rev.\  {\bf D76}, 043502 (2007).
  [arXiv:0705.0079 [hep-ph]].

\bibitem{Pilaftsis:2003gt}
  A.~Pilaftsis, T.~E.~J.~Underwood,
  Nucl.\ Phys.\  {\bf B692}, 303-345 (2004).
  [arXiv:hep-ph/0309342 [hep-ph]].

\bibitem{Conlon:2007gk}
  J.~P.~Conlon, F.~Quevedo,
  JCAP {\bf 0708}, 019 (2007).
  [arXiv:0705.3460 [hep-ph]].

\bibitem{Bassett:2005xm}
  B.~A.~Bassett, S.~Tsujikawa, D.~Wands,
  Rev.\ Mod.\ Phys.\  {\bf 78}, 537-589 (2006).
  [astro-ph/0507632].

\bibitem{Asaka:2000ew}
  T.~Asaka, T.~Yanagida,
  Phys.\ Lett.\  {\bf B494}, 297-301 (2000).
  [hep-ph/0006211].

\bibitem{Anderson:1991zb}
  G.~W.~Anderson, L.~J.~Hall,
  Phys.\ Rev.\  {\bf D45}, 2685-2698 (1992).


\bibitem{Coleman:1977py}
  S.~R.~Coleman,
  Phys.\ Rev.\  {\bf D15}, 2929-2936 (1977).

\bibitem{Megevand:2009gh}
  A.~Megevand, A.~D.~Sanchez,
  Nucl.\ Phys.\  {\bf B825}, 151-176 (2010).
  [arXiv:0908.3663 [hep-ph]].


\end{thebibliography}
\end{document}